\pdfoutput=1
\newcommand*{\ATLASLATEXPATH}{}
\documentclass[cernpreprint,texlive=2016,txfonts,UKenglish]{atlasdoc}
\usepackage{\ATLASLATEXPATH atlaspackage}
\usepackage{\ATLASLATEXPATH atlasbiblatex}

\usepackage{\ATLASLATEXPATH atlasphysics}
\graphicspath{{logos/}{figures/}}

\usepackage{mono_paper_2017-defs}

\usepackage{tabularx}
\usepackage{tabulary}

\AtlasTitle{Search for dark matter and other new phenomena in events with an energetic jet and large missing transverse 
momentum using the ATLAS detector}

\author{The ATLAS Collaboration}

\AtlasRefCode{EXOT-2016-27}

\AtlasJournalRef{JHEP 01 (2018) 126}
\AtlasDOI{10.1007/JHEP01(2018)126}

\PreprintIdNumber{CERN-PH-2017-230}

\AtlasJournal{JHEP}

\AtlasAbstract{
Results of a search for new phenomena in final states with an 
energetic jet and large missing transverse momentum are reported.  
The search uses proton--proton collision data corresponding to an 
integrated luminosity of $36.1\,\ifb$ at 
a centre-of-mass energy of $13~\TeV$ collected 
in 2015 and 2016 with the ATLAS detector at the Large Hadron Collider. 
Events are required to have at least one jet with a transverse 
momentum above $250~\GeV$ and no leptons ($e$ or $\mu$).  Several signal regions are 
considered with increasing 
requirements on the missing transverse momentum
above $250~\GeV$. 
Good agreement is observed between the number of events in data and 
Standard Model predictions.
The results are translated into exclusion limits in models with 
pair-produced weakly interacting dark-matter candidates, large extra 
spatial dimensions, and supersymmetric particles in several 
compressed scenarios.
\\
\\
}

\hypersetup{pdftitle={ATLAS document},pdfauthor={The ATLAS Collaboration}}

\begin{document}

\maketitle

% \tableofcontents

%------------------------------------------------------------------------------- 
% INTRODUCTION
%-------------------------------------------------------------------------------
\section{Introduction}
\label{sec:intro}

This paper
presents the results of a search for events containing an energetic 
jet and large missing transverse momentum $\ptmi$ (with magnitude $\met$) in a 
data sample corresponding to a total integrated luminosity of $36.1\,\ifb$.  
The data were collected by the ATLAS Collaboration at the Large Hadron Collider (LHC) from  
proton--proton collisions at a centre-of-mass energy $(\!\!\sqrt{s})$ of 13~\TeV{}. 
The final-state monojet signature 
of at least one energetic jet, $\met>250~\gev$, and no leptons ($e$ or $\mu$)
constitutes a distinctive signature for new physics beyond the Standard Model 
(SM) at colliders.  
The monojet signature has been extensively studied at the LHC in the context of 
searches for large extra spatial dimensions (LED), supersymmetry (SUSY), and 
weakly interacting massive particles (WIMPs) as candidates for dark matter (DM)~\cite{Aaboud:2016tnv,Sirunyan:2017hci,Aaboud:2017buf}.
The results of the analysis are therefore interpreted in terms of each of these 
models, which are described in the following paragraphs.  

A range of astrophysical measurements, such as the rotational speed of stars in galaxies 
and gravitational lensing, point to the existence of a non-baryonic form of 
matter \cite{Trimble:1987ee,Bertone:2004pz,Feng:2010gw}.  The existence of a new, 
weakly interacting massive particle is often hypothesized~\cite{Steigman:1984ac}, 
as it leads to the correct relic density for non-relativistic matter in the early 
universe~\cite{Kolb:1990vq} as measured 
from data from the
Planck~\cite{Adam:2015rua} and 
WMAP~\cite{2013ApJS..208...19H} Collaborations, if the mass is between a few \GeV{} 
and one \TeV{} and if it has electroweak-scale interaction cross sections. 
WIMPs may be pair-produced at the LHC 
and when 
accompanied by 
a jet of particles, for example from initial-state radiation (ISR), these 
events produce the signature of a jet and missing transverse momentum.

As with the initial results obtained in this search channel at 
$\sqrt{s}=13~\TeV$~\cite{Aaboud:2016tnv}, simplified models are used to 
interpret the results, providing a framework to characterize the new 
particles acting as mediators of the interaction between the SM and the 
dark sector~\cite{Abdallah:2015ter,Abercrombie:2015wmb,Buchmueller:2014yoa}. 
The results from simplified models 
involving $s$-channel 
Feynman diagrams such 
as the one shown in Figure~\ref{fig:feynman}\protect\subref{fig1a}  
are comparable to those previously 
obtained~\cite{Aad:2015zva}
by using an effective-field-theory 
approach~\cite{Goodman:2010ku} when the mediator mass considered is 
above $10~\TeV$~\cite{Busoni:2013lha}.  

Results are presented for  DM models where Dirac fermion WIMPs ($\chi$) are pair-produced
from quarks via $s$-channel exchange of a spin-1 mediator particle ($Z_A$) with
axial-vector couplings, a spin-1 mediator particle ($Z_V$) with vector couplings, or a spin-0 pseudoscalar ($Z_P$).  
These models are defined by four
free parameters: the WIMP mass ($m_\chi$); the mediator mass ($m_{Z_A}$, $m_{Z_V}$ or $m_{Z_P}$, depending on the model); the
flavour-universal coupling to quarks ($g_q$), where all three quark generations are included; and the coupling of the mediator to
WIMPs ($g_\chi$).  Couplings to other SM particles are not considered.
In each case, a minimal mediator width is defined, as detailed in Refs.~\cite{Buchmueller:2014yoa,Abercrombie:2015wmb}, which 
in the case of the axial-vector mediator takes the form:

\begin{equation*}
\Gamma(m_{Z_A})_{\textrm{min}}=
\frac{g_\chi^2 m_{Z_A}}{12\pi} \beta_\chi^3 \theta(m_{Z_A}-2m_\chi)
+ \sum_q \frac{3 g_q^2 m_{Z_A}}{12\pi}\beta_q^3 \theta(m_{Z_A}-2m_q)\;,
\end{equation*}
\noindent
where $\theta(x)$ denotes the Heaviside step function and 
$\beta_f=\sqrt{1 - {4m_f^2}/{m_{Z_A}^2}}$ is the velocity in the mediator rest frame of 
fermion $f$ 
(either $\chi$ or $q$) with mass $m_f$.  The quark sum runs over all 
flavours.  The monojet signature in this model emerges from initial-state 
radiation of a gluon as shown in Figure~\ref{fig:feynman}\protect\subref{fig1a}.

\begin{figure}[htb]
\begin{center}
  \subfloat[][]{
 \includegraphics[width=0.30\linewidth]{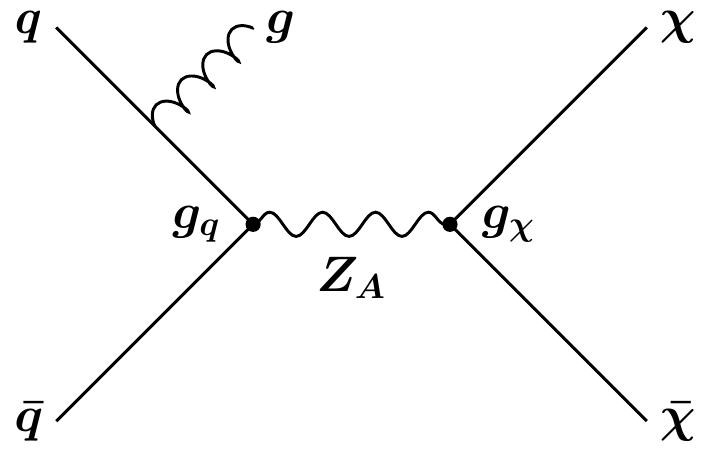}\label{fig1a}
}\\
  \subfloat[][]{
 \includegraphics[width=0.25\linewidth]{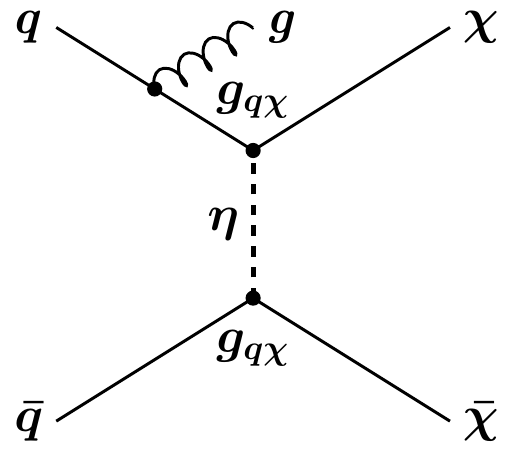}\label{fig1b}
}
  \subfloat[][]{
 \includegraphics[width=0.30\linewidth]{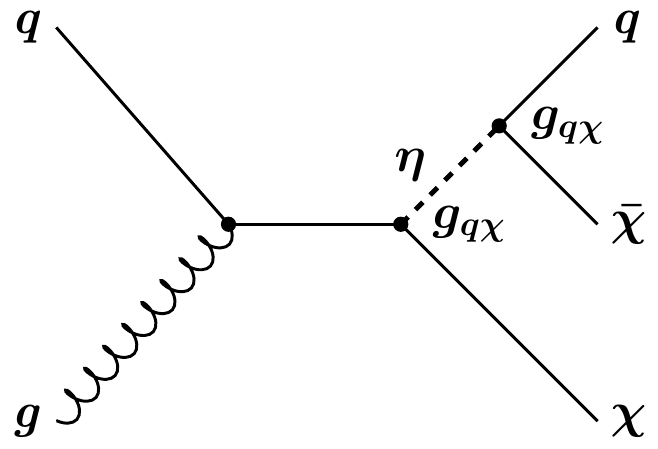}\label{fig1c}
}
  \subfloat[][]{
 \includegraphics[width=0.25\linewidth]{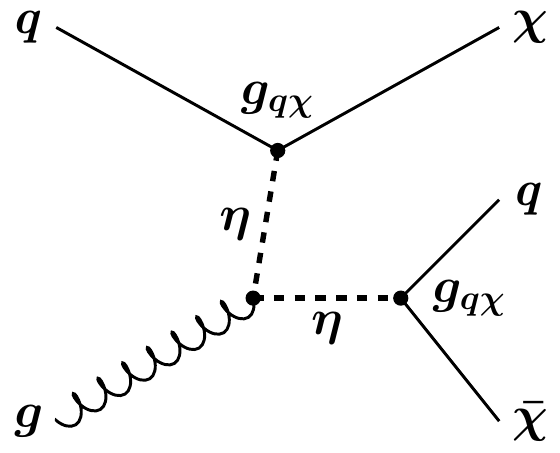}\label{fig1d}
}\\
  \subfloat[][]{
 \includegraphics[width=0.30\linewidth]{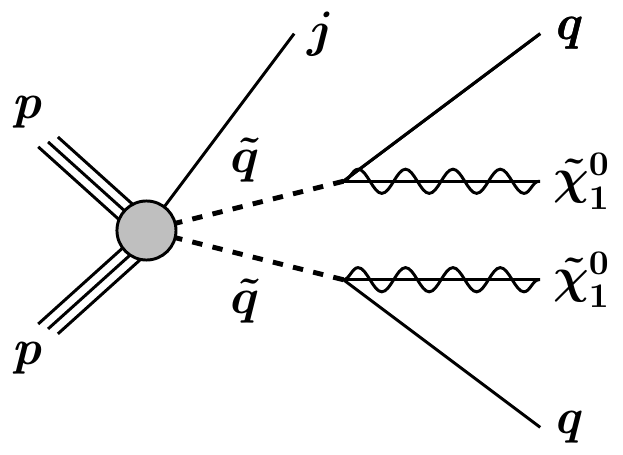}\label{fig1e}
}
\end{center}
\vspace{-.4cm}
\caption{
\protect\subref{fig1a} Diagram for the pair-production of weakly interacting massive particles $\chi$,
with a mediator $Z_A$ with axial-vector couplings exchanged
in the $s$-channel. \protect\subref{fig1b}\protect\subref{fig1c}\protect\subref{fig1d} Example of diagrams for the 
pair-production of weakly interacting massive particles $\chi$ via a coloured scalar mediator $\eta$. 
\protect\subref{fig1e}
A generic diagram for the pair-production of squarks
with the decay mode $\tilde{q} \to q  + \ninoone$.
The presence of a gluon from initial-state
radiation resulting in a jet is indicated for illustration purposes.
}
\label{fig:feynman}
\end{figure}

Results are also presented for a DM model in which  WIMPs are produced via the exchange of a coloured
scalar mediator, which is assumed to couple as a colour-triplet, $\mathrm{SU}(2)$ doublet
to the left-handed quarks \cite{Bell:2012rg, Papucci:2014iwa, Brennan:2016xjh}. 
The model 
contains a variety of new production mechanisms 
such as the
production of 
WIMP pairs via $u$- and $t$-channel diagrams with direct couplings
of dark matter and SM particles or even 
$s$-channel exchange of two mediators,  
leading to a different phenomenology.  A set of representative diagrams relevant for a monojet final state  
are collected in Figures~\ref{fig:feynman}\protect\subref{fig1b}--\ref{fig:feynman}\protect\subref{fig1d}. 
A model with simplified assumptions is defined by the following 
three parameters: $m_\chi$, a single mediator mass ($m_\eta$), and 
a flavour-universal coupling to quarks and WIMPs ($g_{q\chi} \equiv g$).
The mediator is also assumed to couple only to the first two generations of quarks, with minimal decay widths of the form:

 \begin{equation*}
\label{eqn:TSD_width}
\Gamma(\eta)_{\textrm{min}} = \frac{g^{2}}{16\pi m_{\eta}^{3}}\left( m_\eta^{2} 
- m_{q}^{2} - m_\chi^{2} \right)\sqrt{\left( m_\eta^{2} - \left(m_{q} + m_\chi 
\right)^{2}\right)\left( m_\eta^{2} - \left(m_{q} - m_\chi\right)^{2}\right)}, 
\end{equation*}

\noindent
where, to ensure that the DM particle is stable and the mediator width is always 
defined, $m_\chi^{2} + m_{q}^{2} <  m_\eta^{2}$ and %is required and also 
$4m_\chi^{2}/m_\eta^{2} <  \left(1 - m_{q}^{2}/m_\eta^{2} + m_\chi^{2}/m_\eta^{2} \right)^{2}$ are required.

Supersymmetry is a theory of  physics beyond the SM which naturally 
solves the hierarchy problem and provides candidates for dark 
matter~\cite{Miyazawa:1966,Ramond:1971gb,Golfand:1971iw,Neveu:1971rx,Neveu:1971iv,Gervais:1971ji,Volkov:1973ix,Wess:1973kz,Wess:1974tw}. SUSY introduces a new supersymmetric partner (sparticle) 
for each particle in the SM. Specifically, a new scalar field is 
associated with each left- or right-handed quark state.  Two squark 
mass eigenstates $\tilde{q}_1$ and $\tilde{q}_2$ result from the 
mixing of the scalar fields for a particular flavour.  Naturalness 
arguments suggest that the third-generation squarks should be light, 
with masses below about $1~\TeV$~\cite{Barbieri:1987fn}.  In addition, many 
SUSY scenarios have a significant mass difference between the two 
eigenstates in the bottom-squark (sbottom) and top-squark (stop) sectors, 
which leads to light sbottom $\tilde{b}_1$ and stop $\tilde{t}_1$ masses. 
In supersymmetric extensions of the SM that assume R-parity 
conservation~\cite{Fayet:1976et,Fayet:1977yc,Farrar:1978xj,Fayet:1979sa,Dimopoulos:1981zb}, 
sparticles are produced in pairs and the lightest supersymmetric particle 
(LSP) is stable.  
The LSP is assumed to be the lightest 
neutralino $\ninoone$.  

The results 
are interpreted in terms of 
searches for squark production using simplified models in scenarios for 
which the mass difference $\Delta m \equiv m_{\tilde{q}} - m_{\ninoone}$ 
is small (compressed-mass scenario).  Four such scenarios with compressed mass spectra are 
considered: stop-pair production, where the stop decays into a charm quark 
and the LSP ($\tilde{t}_1 \to c + \ninoone$), stop-pair production
in the four-body decay mode
$\tilde{t}_1 \to  b + ff^{'} + \tilde{\chi}^{0}_{1}$, 
sbottom-pair production 
with $\tilde{b}_1 \to b  + \ninoone$, and squark-pair production with 
$\tilde{q} \to q  + \ninoone$  $(q=u,d,c,s)$.  
For relatively small $\Delta m$ ($\lesssim 25~\gev$), both the transverse momenta of the quark 
jets and the \met\ in the final state are small, making it difficult to 
fully reconstruct the signal given the kinematic thresholds for reconstruction.
The presence of jets from ISR is thus used to identify signal events (see 
Figure~\ref{fig:feynman}\protect\subref{fig1e}).  In this case, the squark-pair system is 
boosted, leading to larger $\met$.

The final model considered is that of extra spatial dimensions, the existence 
of which has been postulated to explain the large difference between the 
electroweak unification scale at $O(10^2)~\GeV$ and the Planck scale 
$M_{\mathrm{Pl}}$ at $O(10^{19})~\GeV$.  In the Arkani-Hamed, Dimopoulos, and 
Dvali (ADD) model of LED~\cite{ArkaniHamed:1998rs}, the presence of $n$ 
extra spatial dimensions of size $R$ leads to a fundamental 
Planck scale in $4+n$ dimensions given by ${M_{\mathrm{Pl}}}^2 \sim {M_D}^{2+n}R^n$, 
where $M_D$ is the fundamental scale of the $4+n$-dimensional theory. 
Motivation for the theory comes from the possibility that  $M_D$ is of order 1~\TeV, a scale accessible at the LHC.  
In this model, SM particles and gauge interactions are confined to the 
usual 3+1 space-time dimensions, whereas gravity is free to propagate 
through the entire multidimensional space, which effectively dilutes 
its perceived strength.
The extra spatial dimensions are compactified, resulting in a 
Kaluza--Klein tower of massive graviton modes (KK graviton).  If produced in high-energy 
proton--proton collisions, a KK graviton escaping into the extra dimensions can
be inferred from $\met$, and can lead to a monojet event signature.

The paper 
is organized as follows. The ATLAS detector is described in the 
next Section.  Section~\ref{sec:mc} provides details of the Monte Carlo simulations 
used in the analysis for background and signal processes.  
Section~\ref{sec:recons} discusses the reconstruction 
and identification 
of jets, leptons, 
and missing transverse momentum, while Section~\ref{sec:evt} describes the 
event selection.  The estimation of background contributions and the 
study of systematic uncertainties are discussed in Sections~\ref{sec:backg} 
and~\ref{sec:syst}.  The results are presented in Section~\ref{sec:results} 
and are interpreted in terms of limits in models of WIMP-pair production, 
ADD, and SUSY in compressed scenarios.
Finally, Section~\ref{sec:sum} is devoted to the conclusions.

% ------------------------------------------------------------------------------
% ATLAS DETECTOR
%-------------------------------------------------------------------------------
\section{ATLAS detector}
\label{sec:atlas}

The ATLAS detector~\cite{Aad:2008zzm} covers almost the whole solid 
angle\footnote{ATLAS uses a right-handed coordinate system 
with its origin at the nominal interaction point (IP) in the centre 
of the detector and the $z$-axis along the beam pipe. The $x$-axis 
points from the IP to the centre of the LHC ring, and the $y$-axis 
points upward. Cylindrical coordinates $(r,\phi)$ are used in the 
transverse plane, $\phi$ being the azimuthal angle around the 
$z$-axis. The pseudorapidity is defined in terms of the polar angle 
$\theta$ as $\eta=-\ln\tan(\theta/2)$.}
around the collision point with layers of tracking detectors, calorimeters and muon
chambers. The ATLAS inner detector  
 covers the pseudorapidity range $|\eta|<2.5$. 
It consists of a silicon pixel detector, a silicon microstrip detector, and a straw-tube tracker that also measures transition radiation for particle identification, all immersed in a 2 T axial  magnetic field produced by a solenoid.
During the first LHC long shutdown, a new tracking layer, known as the insertable B-Layer~\cite{Capeans:1291633}, 
was added just outside a narrower beam pipe at a radius of 33 mm.

High-granularity lead/liquid-argon (LAr) electromagnetic sampling calorimeters cover the pseudorapidity
range $|\eta|<$~3.2. Hadronic calorimetry in the range $|\eta|<$~1.7 is provided by a steel/scintillator-tile calorimeter, consisting of a large barrel and two smaller extended barrel cylinders, one on either side of
the central barrel. In the endcaps ($|\eta|>$~1.5), copper/LAr and tungsten/LAr hadronic
calorimeters match the outer $|\eta|$ limits of the endcap electromagnetic calorimeters. The LAr
forward calorimeters provide both the electromagnetic and hadronic energy measurements, and extend
the coverage to $|\eta| < 4.9$.

 The muon spectrometer measures the deflection of muons 
in the magnetic field provided by
large superconducting air-core toroidal magnets
in the pseudorapidity range  $|\eta|<2.7$, instrumented with separate trigger and high-precision tracking chambers. 
Over most of the $\eta$ range, a measurement of the track coordinates in the bending direction of the magnetic 
field is provided
 by monitored drift tubes. Cathode strip chambers  with higher granularity are used in the innermost plane over $2.0 < |\eta| < 2.7$.
The muon    
fast trigger detectors cover the pseudorapidity range $|\eta| < 2.4$ and provide a measurement of the coordinate in the non-bending plane.

The data were collected using an online  two-level trigger system~\cite{Aaboud:2016leb}  that selects events  of interest and reduces 
the event rate from an average of 33 MHz to about 1\,kHz for recording and offline processing.

% -------------------------------------------------------------------
% MC Simulation
% -------------------------------------------------------------------
\section{Monte Carlo simulation}
\label{sec:mc}
\label{sec:mc}

Monte Carlo (MC) simulated event samples are used to compute detector 
acceptance and reconstruction efficiencies, determine signal and 
background contributions, and estimate systematic uncertainties in the 
final results.  Samples are processed with the full ATLAS detector 
simulation~\cite{Aad:2010ah} based on {\normalfont \scshape Geant4}~\cite{Agostinelli:2002hh}.   
Simulated events are then reconstructed and analysed with the same 
analysis chain as for the data, using the same trigger and event 
selection criteria.  The effects of multiple proton--proton 
interactions in the same or neighbouring bunch-crossings (pile-up) are taken into 
account by overlaying 
simulated minimum-bias events 
from {\PYTHIA}~8.205~\cite{pythia} 
onto the hard-scattering process, 
distributed according to the frequency in data.

% --------------------------
% SIGNAL SAMPLES 
% --------------------------
\subsection{Signal simulation}

WIMP $s$-channel signal samples are simulated in {\normalfont \scshape Powheg-Box} 
v2~\cite{Alioli:2010xd,Frixione:2007vw,Nason:2004rx} (revision 3049) 
using two implementations of simplified models, introduced in 
Ref.~\cite{Haisch:2013ata}.  The DMV model of WIMP-pair production is 
used for $s$-channel spin-1 axial-vector or vector mediator exchange at 
next-to-leading order (NLO) in the strong coupling, and the {DMS\_tloop} model is used for 
WIMP-pair production with the $s$-channel spin-0 pseudoscalar mediator 
exchange with the full quark-loop calculation at leading order (LO)~\cite{Haisch:2015ioa}.  
Renormalization and 
factorization scales are set to $H_\mathrm{T}/2$ on an event-by-event 
basis, where $H_\mathrm{T}=\sqrt{m_{\chi\chi}^2+p_{\mathrm{T}, 
j1}^2}+p_{\mathrm{T}, j1}$ is defined by the invariant mass of the WIMP 
pair ($m_{\chi\chi}$) and the transverse momentum of the 
highest-$p_{\mathrm{T}}$ parton-level jet   
($p_{\mathrm{T}, j1}$).  The mediator propagator is described by a 
Breit--Wigner distribution. 
Events are generated using the NNPDF30~\cite{Ball:2014uwa} parton 
distribution functions (PDFs) and interfaced to {\PYTHIA}~8.205 with 
the A14 set of tuned parameters (tune)~\cite{tune14} for parton showering, 
hadronization and the underlying event.  Couplings of the 
mediator to WIMP particles and those of the SM quarks are set to $g_\chi=1$ and 
$g_q=1/4$ for the DMV model whereas both couplings are set to one in the 
case of the DMS\_tloop model.  A grid of samples is produced for WIMP 
masses ranging from $1~\GeV$ to $1~\TeV$ and mediator masses between 
$10~\GeV$ and $10~\TeV$.

Samples for DM production in the coloured scalar mediator model are 
generated with MG5\_aMC@NLO v2.3.3~\cite{Alwall:2014hca} at LO using  NNPDF23LO~\cite{Ball:2012cx} PDFs and interfaced to {\PYTHIA} 8.186 with the A14 tune
for modelling of parton showering, hadronization and the underlying event.
The generation of the different subprocesses is performed following a
procedure outlined in Ref.~\cite{Papucci:2014iwa}. 
Specifically, 
the generation is split 
between DM production with an off-shell mediator and on-shell mediator production followed by decay, 
and the associated production of up to two
partons in the final state is included.  As already mentioned, only diagrams involving the first two
quark generations are  considered and processes with electroweak bosons
are suppressed.
The matching between \MADGRAPH and \PYTHIA is                                                                                                                              
performed following the CKKW-L prescription~\cite{Lonnblad:2011xx}. 
The parton matching scale is set to $m_\eta$/8, where $m_\eta$ denotes the mass of the mediator,   
in the case of 
mediator-pair production,
and to $30~\GeV$ otherwise. 
This particular choice of matching scales optimizes the generation of the samples in the full phase space, 
and minimizes the impact from scale variations on the shape of the predicted 
kinematic distributions.  
The coupling is set to $g= 1$, and a grid of 
samples is produced for WIMP masses ranging from $1~\GeV$ to $1~\TeV$ and mediator masses between
$100~\GeV$ and $2.5~\TeV$.

SUSY signals for stop-pair production 
are generated with MG5\_aMC@NLO v2.2.3   
and interfaced to {\PYTHIA} 8.186 with the A14 tune for modelling of 
the squark decay, parton showering, hadronization, and the underlying 
event. The PDF set used for the generation is NNPDF23LO, and the 
renormalization and factorization scales are set to 
$\mu = \sum_i \sqrt{m_i^2 + p_{T,i}^2}$, where the sum runs
over all final-state particles from the hard-scatter process.
The matrix-element calculation is performed at tree level, and 
includes the emission of up to two 
additional partons. Matching 
to parton-shower calculations is accomplished by the 
CKKW-L prescription, with a matching scale 
set to one quarter of the pair-produced superpartner mass. 
Signal cross sections are calculated at NLO in the strong coupling 
constant, adding the resummation of soft-gluon emission at 
next-to-leading-logarithm (NLO+NLL) accuracy~\cite{Beenakker:1997ut,Beenakker:2010nq,Beenakker:2011fu}. The
nominal cross section and its uncertainty are taken from an envelope
of cross-section predictions using different PDF sets and 
factorization and renormalization scales, as described in
Ref.~\cite{Borschensky:2014cia}.
Simulated samples are produced with squark masses in the range between 
$250~\GeV$ and $700~\GeV$, and squark--neutralino mass differences $\Delta m$ varying between $5~\GeV$ and 
$25~\GeV$.

Simulated samples for the ADD LED model with different numbers of extra 
dimensions in the range $n = 2$--$6$ and a fundamental scale $M_D$ in the range $3.0$--$5.3~\TeV$ 
are generated using {\PYTHIA}~8.205 with NNPDF23LO  
PDFs. The renormalization scale is set to the geometric mean of the squared 
transverse masses of the two produced particles, $\sqrt{(p_{\mathrm{T},G}^2 
+ m_G^2)(p_{\mathrm{T},p}^2 + m_p^2)}$, where  $p_{\mathrm{T},G}$ and $m_G$  
($p_{\mathrm{T},p}$ and $m_p$) denote, respectively, the mass and the 
transverse momentum of the KK graviton (parton) in the final state. The 
factorization scale is set to the minimum transverse mass, $\sqrt{\pt^2 + 
m^2}$, of the KK graviton and the parton.

% ===================================
% BACKGROUND SAMPLES
% ===================================
\subsection{Background simulation}

After applying the selection described in Section~\ref{sec:evt},
the primary SM background contributing to monojet event signatures is 
$\znn$+jets.  There are also significant contributions from $W$+jets 
events, primarily from $\wtn$+jets.  Small contributions are expected 
from $\zll$+jets ($\ell = e, \mu, \tau$), multijet, $t\bar{t}$, 
single-top, and diboson ($WW,WZ,ZZ$) processes. Contributions from top-quark  
production associated with additional vector bosons ($\ttb + W$, $\ttb 
+ Z$, or $t + Z + q/b$ processes) are negligible and not considered in this analysis.

% ---W/Z+jets 

Events containing $W$ or $Z$ bosons with associated jets are simulated 
using the {\SHERPA}~2.2.1~\cite{sherpa} event generator. Matrix elements (ME) 
are calculated for up to two partons at NLO  
and four partons at LO using 
OpenLoops~\cite{Cascioli:2011va} and  
Comix~\cite{Gleisberg:2008fv},   
and merged with the {\SHERPA} parton shower 
(PS)~\cite{Schumann:2007mg} using the ME+PS@NLO 
prescription~\cite{Hoeche:2012yf}. The 
NNPDF3.0NNLO~\cite{Ball:2014uwa} PDF 
set is used in conjunction with a dedicated 
parton-shower tuning developed by the authors of \SHERPA. The MC 
predictions are initially normalized to next-to-next-to-leading-order 
(NNLO) perturbative QCD (pQCD) predictions according to 
DYNNLO~\cite{Catani:2009sm,Catani:2007vq} using the  MSTW2008 90$\%$ CL 
NNLO PDF set~\cite{mstw}.

The $W$+jets and $Z$+jets MC predictions are reweighted to account for 
higher-order QCD and electroweak corrections as described in 
Ref.~\cite{Lindert:2017olm}, where parton-level predictions for 
$W/Z$+jets production, including 
 NLO QCD corrections and 
NLO electroweak corrections supplemented by Sudakov logarithms at two 
loops, are provided as a function of the vector-boson $\pt$, improving 
the description of  the measured $Z$-boson $\pt$ 
distribution~\cite{Aaboud:2017hbk}.  The predictions are provided 
separately for the different $W$+jets and $Z$+jets processes together 
with the means for a proper estimation of theoretical uncertainties 
and their correlations (see Section~\ref{sec:syst}). The reweighting 
procedure takes into account the difference between the QCD NLO 
predictions as included already in {\SHERPA} and as provided by the 
parton-level calculations.

For the generation of $\ttb$ and single top quarks in the $Wt$-channel 
and $s$-channel, the {\normalfont \scshape Powheg-Box} v2~\cite{powheg} event generator is used with 
CT10~\cite{ct10} PDFs. Electroweak $t$-channel single-top-quark events are generated 
using the  {\normalfont \scshape Powheg-Box} v1 event generator. This event generator uses the four-flavour 
scheme to calculate NLO matrix elements, with the CT10 four-flavour 
PDF set. The parton shower, hadronization, and underlying event are 
simulated using {\PYTHIA}~8.205 with
the A14 tune.
The top-quark mass is set to $172.5~\GeV$.
The EvtGen v1.2.0 program~\cite{EvtGen} is used to model the decays of
the bottom and charm hadrons.
Alternative samples are generated using \MGMCatNLO~(v2.2.1)~\cite{Alwall:2014hca} 
interfaced to \HERWIGpp~(v2.7.1)~\cite{Bahr:2008pv}
in order to estimate the effects of the choice of matrix-element event generator and parton-shower algorithm.

Diboson samples ($WW$, $WZ$, 
and $ZZ$ production) are generated using either {\SHERPA}~2.2.1 or {\SHERPA}~2.1.1  with NNPDF3.0NNLO or CT10nlo PDFs, respectively,  
and are normalized to NLO pQCD predictions~\cite{Campbell:2011bn}.  
Diboson samples are also generated using {\normalfont \scshape Powheg-Box}~\cite{Frixione:2007vw} interfaced to 
{\PYTHIA}~8.186 and using CT10 PDFs for studies of systematic uncertainties.

% ====================
% RECONSTRUCTION OF PHYSICS OBJECTS
% ====================
\section{Event reconstruction}
\label{sec:recons}

Jets are reconstructed from energy deposits in the calorimeters using the $\akt$ 
jet algorithm~\cite{paper:antikt,Cacciari:2011ma} with the radius parameter (in $y$--$\phi$ space) 
set to 0.4. The measured jet transverse momentum is corrected for detector effects
by weighting energy 
deposits arising from electromagnetic and hadronic showers differently.  In 
addition, jets are corrected for contributions from pile-up, as described in 
Ref.~\cite{Aad:2011he}.  Jets with $p_{\mathrm T} > 20~\GeV$ and $|\eta| 
<2.8$ are considered in the analysis. Track-based variables to suppress 
pile-up jets have been developed, and a combination of two such variables, called the 
jet-vertex tagger~\cite{ATLAS-CONF-2014-018}, is constructed. In order to remove jets originating from 
pile-up collisions, for central jets ($|\eta| < 2.4$) with $p_{\mathrm T} < 50~\GeV$ 
a significant fraction of the tracks associated with each jet must have an origin 
compatible with the primary vertex, as defined by the jet-vertex tagger.

Jets with $\pt > 30~\GeV$ and $|\eta|<2.5$ are identified as $b$-jets if tagged by a multivariate algorithm which uses 
information about the impact parameters of inner-detector tracks matched to the 
jet, the presence of displaced secondary vertices, and the reconstructed flight 
paths of $b$- and $c$-hadrons inside the jet~\cite{Aad:2015ydr,ATL-PHYS-PUB-2016-012}. A 60$\%$ efficient 
$b$-tagging working point, as determined in a simulated sample of $\ttbar$ events, 
is chosen.  This corresponds to a rejection factor of approximately  1500, 35 
and 180 for light-quark and gluon jets, $c$-jets, and $\tau$-leptons decaying 
hadronically, respectively.  

The presence of 
electrons or muons in the final state is used 
in the analysis to define control samples and to reject background contributions in 
the signal regions (see Sections~\ref{sec:evt} and~\ref{sec:backg}). 

Electrons are found by combining energy deposits in the calorimeter
with tracks found in the inner detector, and are initially required to 
have $p_{\mathrm T} > 20~\GeV$ and $|\eta| 
<2.47$, 
to satisfy the 
`Loose' electron shower shape and track selection criteria described in 
Refs.~\cite{ATLAS-CONF-2014-032}, and must also be isolated. 
The latter uses track-based isolation requirements with an efficiency of about 99$\%$, as determined using $\zee$ data. 
Overlaps between identified electrons 
and jets with $\pt > 30~\GeV$ in the final state are resolved. Jets are discarded if they are not 
$b$-tagged and their separation $\Delta R = \sqrt{(\Delta \eta)^2 + (\Delta \phi)^2}$ 
from an identified electron is less than~0.2. Otherwise, the electron is removed as it most likely originates from a 
semileptonic $b$-hadron decay. The electrons separated by $\Delta R$ between 0.2 
and  0.4 from any remaining jet are removed.

Muon candidates are 
formed by combining information from the muon spectrometer and
inner tracking detectors. They are 
required to pass 'Medium' identification requirements, as described in Ref.~\cite{Aad:2016jkr}, and   
to have  $p_{\mathrm T} > 10~\GeV$ and  $|\eta| < 2.5$.
Jets with $p_{\mathrm T} > 
30~\GeV$ and fewer than  three tracks with $\pt > 0.5~\GeV$ associated with them are 
discarded if their separation $\Delta R$ from an identified muon is less than~0.4. 
The muon is discarded if it is matched to a jet with 
$p_{\mathrm T} > 30~\GeV$ that has at least three tracks 
associated with it.

The $\met$ value is reconstructed using all energy deposits in the calorimeter up to 
pseudorapidity $|\eta| =  4.9$. Clusters associated with either electrons,  
photons or jets with $\pt >20~\GeV$ 
make use of the corresponding calibrations. 
Softer jets and clusters not associated with electrons, photons or jets 
are calibrated using tracking 
information~\cite{ATL-PHYS-PUB-2015-023}.  As discussed below, in this analysis the  
missing transverse momentum
is not corrected for the presence of muons in the final state.

% ====================
% EVENT SELECTION
% ====================
\section{Event selection}
\label{sec:evt}

The data sample considered corresponds to a total integrated 
luminosity of $36.1\,\ifb$, and was collected in 2015 and 2016.
The uncertainty in the combined 2015+2016 integrated luminosity is 3.2$\%$. 
It is derived, following a methodology similar to that detailed in Ref.~\cite{Aaboud:2016hhf}, 
from a calibration of the luminosity scale using $x$--$y$ beam-separation scans performed in August 2015 and May 2016.
The data were collected using a trigger that selects 
events with $\met$ above $90~\GeV$, as computed from calorimetry information 
at the final stage of the two-level trigger system.  After 
analysis selections, the trigger was measured to be fully efficient for 
events with $\met > 250~\GeV$, as determined using a data sample with 
muons in the final state.  
Events are required to have at least one reconstructed primary vertex 
consistent with the beamspot envelope and that contains at least two 
associated tracks of $\pt>0.4~\GeV$.  When more than one such vertex is 
found, the vertex with the largest summed $\pt^2$ of the associated 
tracks is chosen.  Events having identified muons with $\pt > 10~\GeV$ or 
electrons with $\pt > 20~\GeV$ in the final state are vetoed. 

Events are selected with $\met > 250~\GeV$, 
a leading jet with $p_{\mathrm{T}, j1} > 250~\GeV$ and $|\eta| < 2.4$, and
a maximum of four jets with $\pt > 30~\GeV$ and $|\eta| < 2.8$.
Separation in the azimuthal angle of $\Delta\phi(\mathrm{jet},{\ptmi}) 
> 0.4$ between the missing transverse momentum direction and each 
selected jet is required to reduce the multijet background 
contribution, where a large $\met$ can originate from jet energy 
mismeasurement.

Jet quality criteria~\cite{clean} are imposed, which involve selections based 
on quantities such as the pulse shape of the energy depositions in the 
cells of the calorimeters, electromagnetic fraction in the calorimeter, 
calorimeter sampling fraction, and the charged-particle fraction.\footnote{
The charged-particle fraction is defined as 
$f_{{\mathrm{ch}}}=\sum p_{\mathrm T}^{{\textrm{track,jet}}} 
/p_{\mathrm T}^{{\textrm{jet}}}$, where $\sum p_{\mathrm 
T}^{{\textrm{track,jet}}}$ is the scalar sum of the transverse momenta 
of tracks associated with the primary vertex within a cone of radius 
$\Delta R=0.4$ around the jet axis, and $p_{\mathrm T}^{{ \textrm{jet}}}$ 
is the transverse momentum of the jet as determined from calorimetric measurements.}
Loose selection criteria are applied to all jets with $\ptjet > 30~\GeV$ 
and $|\eta| < 2.8$, which remove anomalous energy depositions due to 
coherent noise and electronic noise bursts in the calorimeter~\cite{Aad:2013zwa}.
Events with any jet not satisfying  the loose criteria, as described in Ref.~\cite{clean}, 
are discarded.

Non-collision backgrounds, for example energy depositions in the calorimeters 
due to muons of beam-induced or cosmic-ray origin, are 
suppressed by imposing tight selection criteria on the leading jet and the ratio 
of the jet charged-particle fraction to the calorimeter sampling 
fraction,\footnote{The variable $f_{\small{\textrm{max}}}$ denotes the maximum fraction of 
the jet energy collected by a single calorimeter layer.} 
$f_{{\textrm{ch}}}/f_{\small{\textrm{max}}}$, is required to be larger than 0.1. 
These requirements have a negligible effect on the signal efficiency. 

The analysis uses two sets of signal regions, with inclusive and exclusive 
$\met$ selections, where the regions are defined with increasing
$\met$ thresholds from $250~\GeV$ to $1000~\GeV$ (Table~\ref{tab:sr}).
The inclusive selections are used for a model-independent search for new physics, and
the exclusive selections are used for the interpretation of the results
within different models of  new physics. 

% 
% ================================
% SUMMARY TABLE
% ================================

\begin{table*}[!ht]
\renewcommand{\baselinestretch}{1}
\caption{Inclusive (IM1--IM10) and exclusive (EM1--EM10) signal regions 
with increasing $\met$ thresholds from $250~\GeV$ to $1000~\GeV$.
In the case of IM10 and EM10, both signal regions contain the same 
selected events in data. In the case of the IM10 signal region, the background 
predictions are computed considering only data and simulated events with $\met > 1~\TeV$, whereas 
the EM10 background prediction is obtained from fitting the full $\met$ shape in data and simulation,  
as described in Section~\ref{sec:backg}.}
\begin{center}
\begin{footnotesize}
\begin{tabular*}{\textwidth}{@{\extracolsep{\fill}}lccccccccccc}\hline
Inclusive (IM)& IM1 & IM2  & IM3      & IM4      & IM5      & IM6      & IM7      & IM8      & IM9      & IM10    \\
$\met$ [\GeV] & $>\!250$ & $>\!300$ & $>\!350$ & $>\!400$ & $>\!500$ & $>\!600$ & $>\!700$ & $>\!800$ & $>\!900$ & $>\!1000$ \\ \hline
Exclusive (EM)& EM1 & EM2  & EM3      & EM4      & EM5      & EM6      & EM7      & EM8      & EM9      & EM10    \\ 
$\met$ [\GeV] &\!250--300\!&\!300--350\!&\!350--400\!&\!400--500\!&\!500--600\!&\!600--700\!&\!700--800\!&\!800--900\!&\!900--1000\!& $>\!1000$ \\  \hline
\end{tabular*}
\end{footnotesize}
\end{center}
\label{tab:sr}
\end{table*}

% ====================
% BACKGROUND EXPECTATIONS
% ====================
\section{Background estimation}
\label{sec:backg}
The $W$+jets, $Z$+jets, and top-quark-related backgrounds are constrained using 
MC event samples normalized with data in selected control regions. By construction, there is no overlap 
between events in the signal and the  
different control regions. 
The control regions are defined using the same
requirements for $\met$, leading-jet $\pt$, event topologies, and jet vetoes as in the
signal regions,  such that no extrapolation in $\met$ or jet $\pt$ is needed from control to signal regions.
The normalization factors are
extracted simultaneously using a global fit that includes systematic uncertainties,
to properly take into account correlations.

Different control samples  
are used to help constrain the yields of the $W$+jets and $Z$+jets background processes in the signal regions.
This includes $\wmn$+jets, $\wen$+jets, and $\zmm$+jets  control samples, enriched in $\wmn$+jets, $\wen$+jets, and $\zmm$+jets  background processes, respectively.
The dominant $\znn$+jets and $\wtn$+jets background contributions are constrained in the fit by using both  $W$+jets control regions and the $\zmm$+jets control region. 
As discussed in Section~\ref{sec:fitback}, this translates into a reduced uncertainty
in the estimation of the main irreducible background contribution,  due to a partial 
cancelling out of
systematic uncertainties and the
superior  statistical power of the $W$+jets control sample in data,
compared to that of the $\zmm$+jets control sample.
A small  $\zee$+jets and $\ztt$+jets background contribution is also constrained via the $W$+jets and $\zmm$+jets control samples.\footnote{The use of an additional $\zee$+jets control
sample to help constrain the $\zee$+jets and $\znn$+jets background contributions
leads to an insignificant improvement in the background determination~\cite{Aaboud:2016tnv}.}

Finally, 
a top control sample constrains 
top-quark-related background processes. 
The remaining SM backgrounds from diboson processes  
are determined using MC simulated samples, while the multijet background  contribution
is extracted from data.
The  
contributions from non-collision backgrounds are
estimated in data using the beam-induced background identification techniques described in Ref.~\cite{Aad:2013zwa}.

In the following subsections, details 
of the definition of the $W/Z$+jets and top control regions, and of the data-driven determination of
the multijet and beam-induced backgrounds are given. This is followed by a description of the 
background fits.

\subsection{Control samples}

A $\wmn$+jets
control sample is selected by requiring a muon consistent with originating from the primary vertex with $\pt > 10~\GeV$, and  
transverse mass
in the range  $30 <  m_{\mathrm T} <  100~\GeV$. 
The transverse mass $m_{\mathrm T} = \sqrt{2\pt^{\ell}\pt^{\nu}[1-\cos(\phi^{\ell}-\phi^{\nu})]}$  is defined by the lepton  and neutrino transverse momenta,   
where the $(x,y)$ components of the neutrino momentum are taken to be
the same as the corresponding $\ptmi$ components. Events with identified electrons in the final state are vetoed.  
In addition, events with an identified $b$-jet in the final state are vetoed in order to reduce the contamination from top-quark-related processes. 
Similarly, a $\zmm$+jets control
sample is selected by  requiring the presence of two muons with $\pt >10~\GeV$ and 
invariant mass in the range $66 < m_{\mu \mu} < 116~\GeV$. In the $\wmn$+jets and  
$\zmm$+jets control regions,  the $\met$ value is not corrected for the presence of 
the muons in the final state,  
motivated by the fact that these control regions
are used to estimate the $\znn$+jets, $\wmn$+jets and $\zmm$+jets backgrounds  
in the signal regions with no identified muons.         
The $\met$-based online trigger used in the analysis
does not include muon information in the $\met$ calculation. This allows the
collection of $\wmn$+jets and $\zmm$+jets control samples with the same trigger as for the signal regions.

A $\wen$+jets-dominated control sample was collected using  online triggers that 
select events with an electron in the final state. The control sample 
is defined 
with an isolated electron candidate with $\pt >30~\GeV$, $30 <  m_{\mathrm T} <  100~\GeV$,  and 
no additional identified leptons in the final state. 
Electron candidates in the 
transition region between the barrel
and endcaps of the electromagnetic calorimeter, $1.37 < |\eta| < 1.52$,  are excluded.
The $\met$ value is corrected by subtracting the contribution from the electron cluster in the calorimeter.
In this way, the measured $\met$ in the event better reflects the magnitude of the $W$-boson $\pt$ in the final state, which 
is necessary for a proper implementation of the $W$-boson $\pT$ reweighting procedure, as explained in Section~\ref{sec:mc}, 
that accounts for higher-order QCD and electroweak corrections.
In order to 
suppress backgrounds from multijet processes with jets faking high-$\pt$ electrons, the events are required to have $\met/\sqrt{H_{\mathrm T}} > 5~\GeV{}^{1/2}$, where 
in this case $\met$ still includes the contribution from the electron energy deposits in the calorimeter and $H_{\mathrm T}$ denotes the scalar
sum of the $\pt$ of the identified jets in the final state.

Finally, a control sample enriched in $\ttbar$ events is constructed using the same selection criteria as in the case of the $\wmn$+jets but 
requiring that at least one of the jets is $b$-tagged. 

\subsection{Multijet background}
\label{sec:qcdback}

The multijet background with large $\met$ mainly originates 
from the misreconstruction of the energy of a 
jet in the calorimeter and to a lesser extent is due to the presence 
of neutrinos in the final state from heavy-flavour hadron decays.  
In this analysis, the multijet background is determined from data, using the jet smearing method as
described in Ref.~\cite{Aad:2012fqa}, which relies  on the assumption that the $\met$ value of multijet events is dominated by  
fluctuations in the jet response in the detector, which can be measured in the data. 
For the IM1 and EM1 selections, the multijet background constitutes about $0.3\%$ and $0.4\%$ of the total background, respectively, and  
it  is negligible for the other signal regions.

\subsection{Non-collision background}
\label{sec:noncoll}

Remaining non-collision background contributions in the signal regions, mostly from %beam-induced 
muons
originating in the particle cascades due to beam-halo protons intercepting the LHC collimators, are estimated following closely 
the methods set out in Ref.~\cite{Aad:2013zwa}.  In particular, the jet timing, $t_j$, calculated from the 
energy-weighted average of the time of the jet energy deposits, defined with respect to the event time in nominal collisions, is 
used. 
A dedicated region enhanced in beam-induced background, 
defined by inverting the tight jet-quality selection imposed on the leading jet, is used to estimate the 
amount of non-collision background  from the fraction of events with a leading-jet timing $|t_j| >  5$~ns.   
The results indicate an almost negligible contribution from non-collision backgrounds in the signal regions.

% ===============
%  BACKGROUND FITS
% ===============

\subsection{Background fit}
\label{sec:fitback}

The use of control regions to constrain the normalization of the dominant background contributions 
reduces the relatively large  theoretical and experimental 
systematic uncertainties, of the order of  $20\%$--$40\%$,  associated with purely simulation-based 
background predictions in the signal regions.
A complete study of systematic uncertainties is carried out, as detailed in Section~\ref{sec:syst}.
To determine the final uncertainty in the total background,  
all systematic uncertainties are treated as 
Gaussian-distributed
nuisance
parameters in a fit based on the profile
likelihood method~\cite{statforumlimits}, which takes into account correlations among systematic variations.
The likelihood also takes into account cross-contamination between different background sources in the control regions.

The $\met$ distribution is the observable used. A simultaneous background-only likelihood fit to the $\met$ distributions in the $\wmn$+jets, $\wen$+jets,  
$\zmm$+jets, and top control regions is performed to normalize and
constrain the background estimates in the signal regions.
In the analysis, two different fitting strategies are considered, potentially giving slightly different results.  
A binned likelihood fit is performed using simultaneously all the exclusive $\met$ regions EM1--EM10, as described in Section~\ref{sec:evt}.
The fit includes a single floating normalization factor common to all $W$+jets and $Z$+jets processes, 
and a single floating normalization factor 
for top-quark-related processes. The nuisance parameters, implementing the impact of systematic uncertainties, are defined bin-by-bin and 
correlations across $\met$ bins are taken into account.  As a result, the fit exploits the information of the shape of the
$\met$ distribution in constraining the normalization of $W/Z$+jets and top-quark-related background.     
In addition, one-bin likelihood fits are performed separately for each of the inclusive regions IM1--IM10. 
In this case, the two normalization factors for $W/Z$+jets and top-quark-related processes, respectively, and the nuisance parameters related to systematic uncertainties refer to the given $\met$ inclusive region. 
 
The results of the background-only fit in the control regions    
are presented in Table~\ref{tab:fitm1} for the $\met > 250~\GeV$ inclusive selection. 
The $W/Z$+jets background predictions 
receive a multiplicative  normalization factor of $1.27$. Similarly,   
top-quark-related processes receive a normalization factor of $1.06$.
When the binned likelihood fit is performed simultaneously over the different exclusive $\met$ regions, thus including information from the shape of the measured $\met$ distribution,
the normalization  factor of the $W/Z$+jets background predictions remains essentially unchanged, dominated by the low-$\met$ region,  
and that of the top-quark-related processes becomes $1.31$, correlated with a less than $1\sigma$ pull of the top-quark-related uncertainties within 
the fit.

%
% --- TABLES
%

\begin{table}[tbh]
\caption{Data and background predictions in the control regions before and after the fit is performed for the $\met > 250~\GeV$ inclusive selection.
The background predictions include both the statistical and systematic uncertainties.
The individual uncertainties are correlated, and do not necessarily add 
in quadrature to the total background uncertainty. The dash ``\textendash'' denotes negligible background contributions.
}
\begin{center}
\setlength{\tabcolsep}{0.0pc}
{\footnotesize
\begin{tabularx}{\textwidth}{lXr@{\,$\pm$\,}lXr@{\,$\pm$\,}lXr@{\,$\pm$\,}lXr@{\,$\pm$\,}l}
\noalign{\smallskip}\hline\noalign{\smallskip}
{\normalfont \bfseries  $\met > 250~\GeV$ Control Regions}        && \multicolumn{2}{c}{$\wmn$}  && \multicolumn{2}{c}{$\wen$}  &&  \multicolumn{2}{c}{$\zmm$}&& \multicolumn{2}{c}{Top} \\ \noalign{\smallskip}\hline\noalign{\smallskip}
Observed events (36.1~fb${}^{-1}$)&& \multicolumn{2}{c}{$110938$}&& \multicolumn{2}{c}{$68973$} && \multicolumn{2}{c}{$17372$}&& \multicolumn{2}{c}{$9729$}   \\ \noalign{\smallskip}\hline\noalign{\smallskip}
SM prediction (post-fit)          &&  110810 & 350 &&  69030 & 260 &&  17440 & 130 &&  9720 & 130 \\ 
\cmidrule{3-13}
$\wen$              &&     7 & 2  &&54500 & 1000      &&     \multicolumn{2}{c}{\textendash}   &&\multicolumn{2}{c}{$0.2_{-0.2}^{+0.4}$}\\  
$\wmn$              && 94940 & 900&&            7 & 7 &&    32 & 3 &&2160 & 650      \\   
$\wtn$              &&  5860 & 160&&     4110 & 140       &&     3 & 1   && 164 & 40     \\  
$\zee$              &&     \multicolumn{2}{c}{\textendash}  &&        5 & 4         &&     \multicolumn{2}{c}{\textendash}   &&   \multicolumn{2}{c}{\textendash}      \\  
$\zmm$              &&  1774 & 75 &&      0.4 & 0.2       && 16360 & 160 &&  59 & 12     \\      
$\ztt$              &&   277 & 21 &&      212 & 15        &&    16 & 3   &&  12 & 2      \\   
$\znn$              &&    37 & 3  &&      1.8 & 0.3       &&     \multicolumn{2}{c}{\textendash}   &&   6 & 1      \\  
$\ttbar$, single top&&  4700 & 790&&     8200 & 1000      &&   486 & 64  &&7220 & 820    \\    
Diboson             &&  3220 & 230&&     2020 & 160       &&   540 & 39  && 108 & 38     \\    
\noalign{\smallskip}\hline\noalign{\smallskip}
SM prediction from simulation (pre-fit)&& 87500 & 8700&&56600 & 5600  && 14100 & 1400 && 9200 & 2000 \\ 
\cmidrule{3-13}
$\wen$              &&    5 & 1    &&43300 & 4700      &&     \multicolumn{2}{c}{\textendash} && \multicolumn{2}{c}{$0.15_{-0.15}^{+0.41}$}   \\  
$\wmn$              &&73700 & 7900 &&      5 & 5       &&    24 & 3&& 1960 & 580            \\   
$\wtn$              && 4600 & 480  && 3260 & 350       &&  2.2 & 0.5  &&  148 & 37     \\                  
$\zee$              &&    \multicolumn{2}{c}{\textendash}    &&    6 & 5         &&    \multicolumn{2}{c}{\textendash}    &&    \multicolumn{2}{c}{\textendash}      \\                  
$\zmm$              && 1420 & 160  &&  0.5 & 0.2       &&13100 & 1400 &&   53 & 11     \\                  
$\ztt$              &&  226 & 29   &&  175 & 20        &&   13 & 3    &&   10 & 2      \\                  
$\znn$              &&   30 & 4    &&  1.5 & 0.3       &&     \multicolumn{2}{c}{\textendash}   &&    5 & 1      \\                  
$\ttbar$, single top&& 4300 & 1200 && 7800 & 2100      &&   460 & 120 && 6900 & 1800   \\                  
Diboson             && 3180 & 230  && 2050 & 170       &&   541 & 40  &&  128 & 44     \\                  
\noalign{\smallskip}\hline\noalign{\smallskip}                              
\end{tabularx}
}
%%%%
\end{center}
\label{tab:fitm1}
\end{table}

Figures~\ref{fig:cr1} and ~\ref{fig:cr2}
show 
the distributions of the $\met$ and the 
leading-jet $\pt$ in data and MC simulation in the different control regions.
In this case, the MC predictions
include the data-driven normalization factors as 
extracted from the binned likelihood fit to the different exclusive $\met$ bins.
Altogether, the MC  simulation provides a good 
description, within uncertainties, of the shape of the measured distributions in the different control regions.

\begin{figure}[!ht]
\begin{center}
 \subfloat[][]{
  \includegraphics[width=0.5\textwidth]{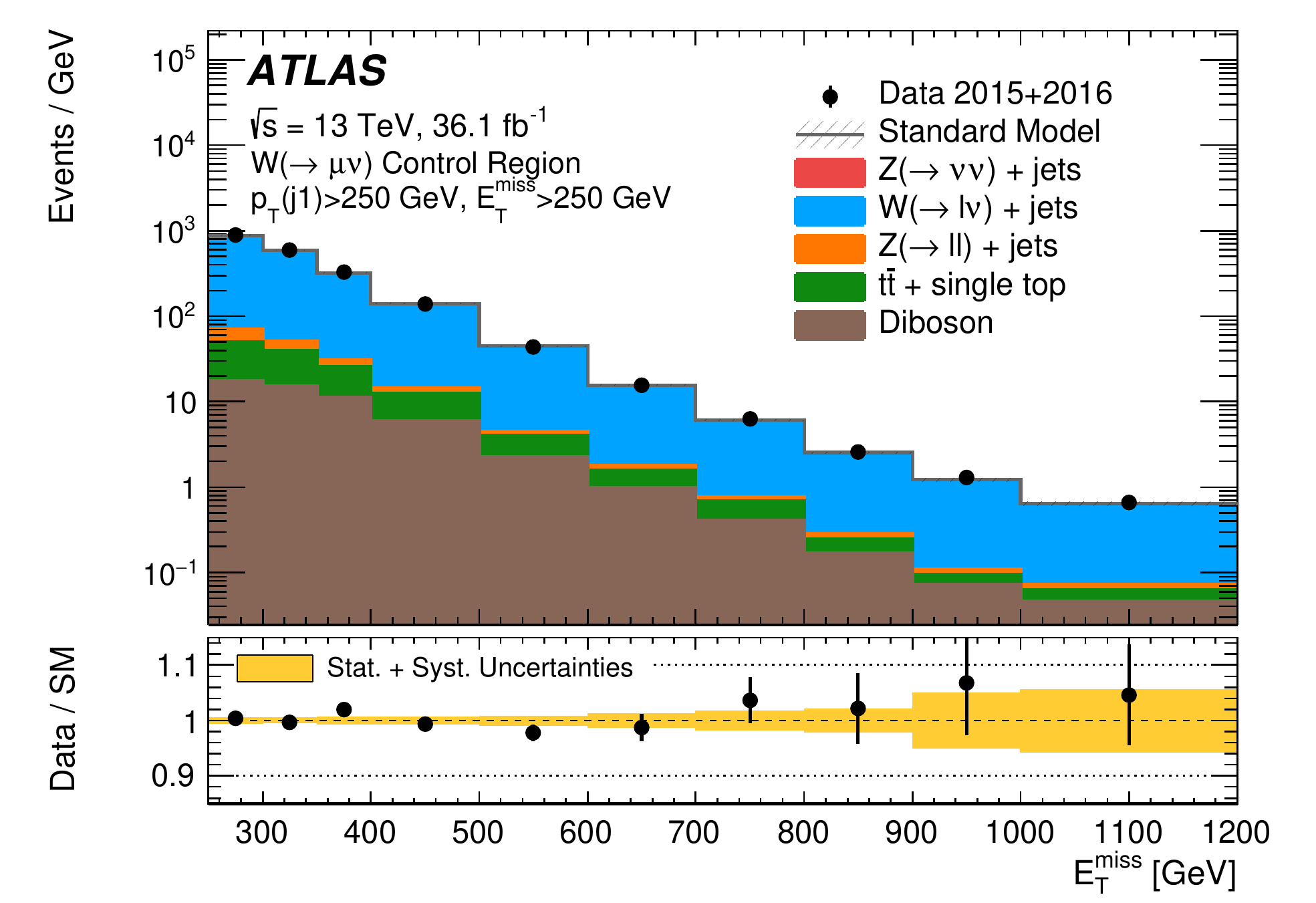}\label{fig2a}
}
 \subfloat[][]{
  \includegraphics[width=0.5\textwidth]{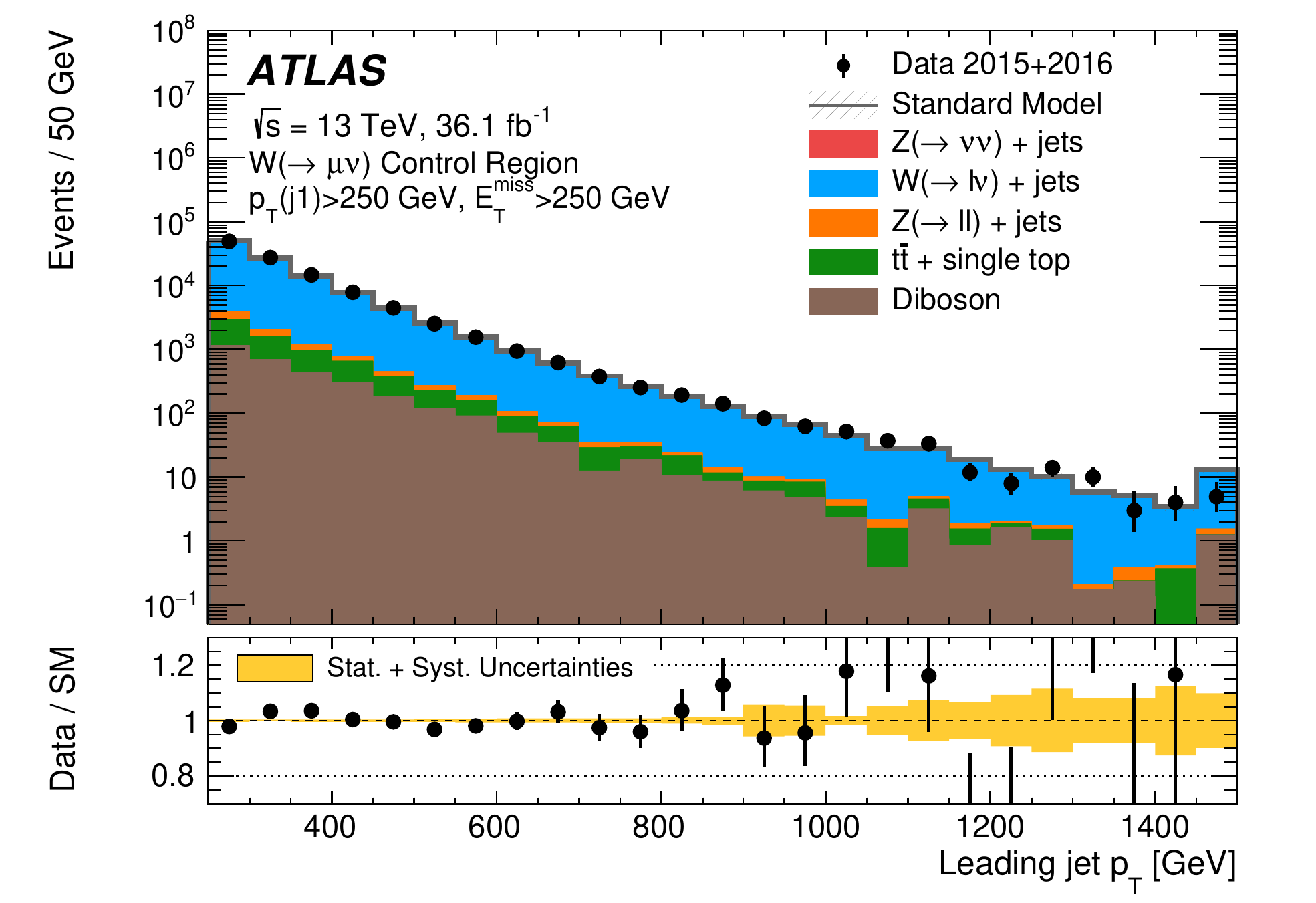}\label{fig2b}
}\\
 \subfloat[][]{
  \includegraphics[width=0.5\textwidth]{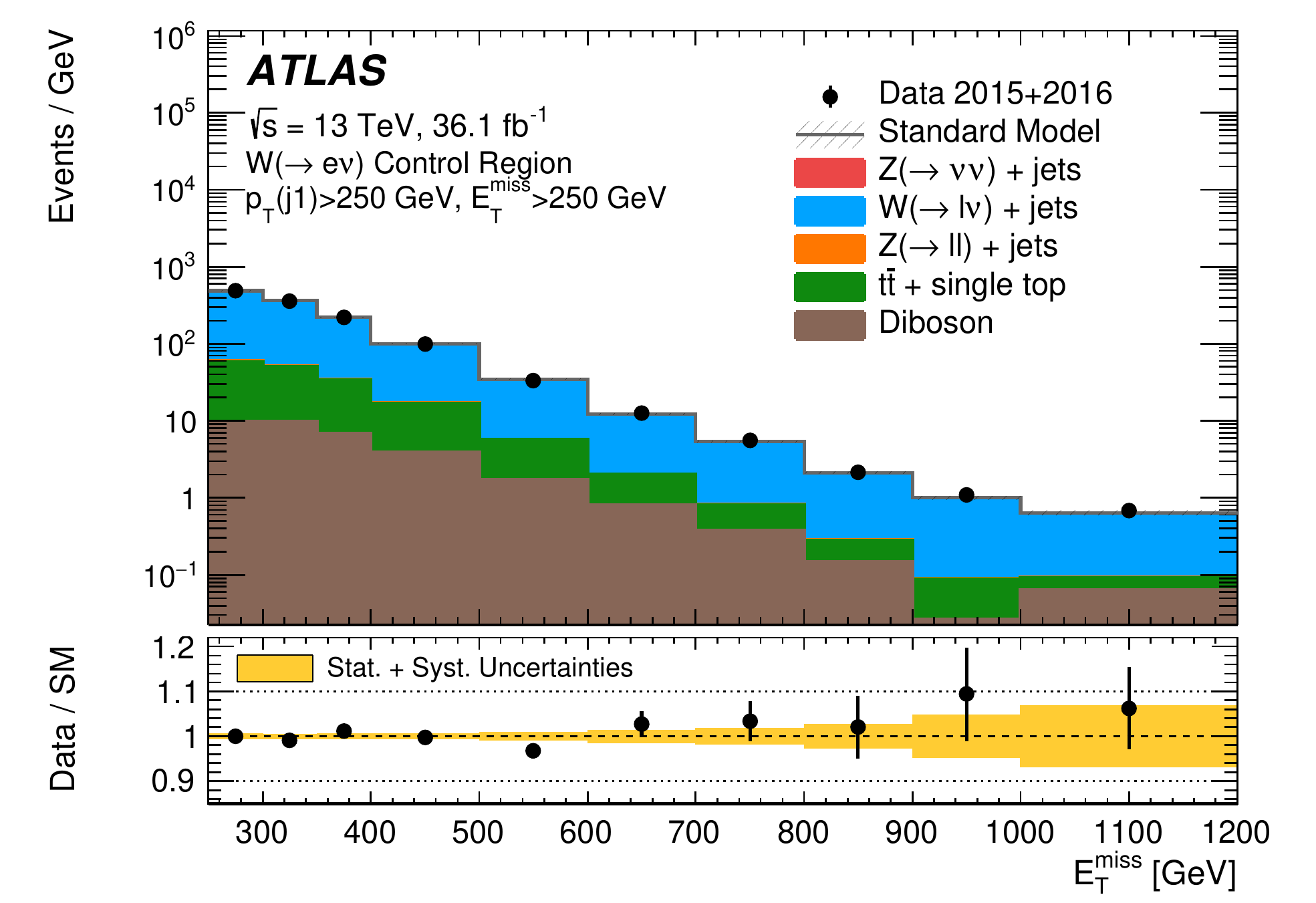}\label{fig2c}
}
 \subfloat[][]{
  \includegraphics[width=0.5\textwidth]{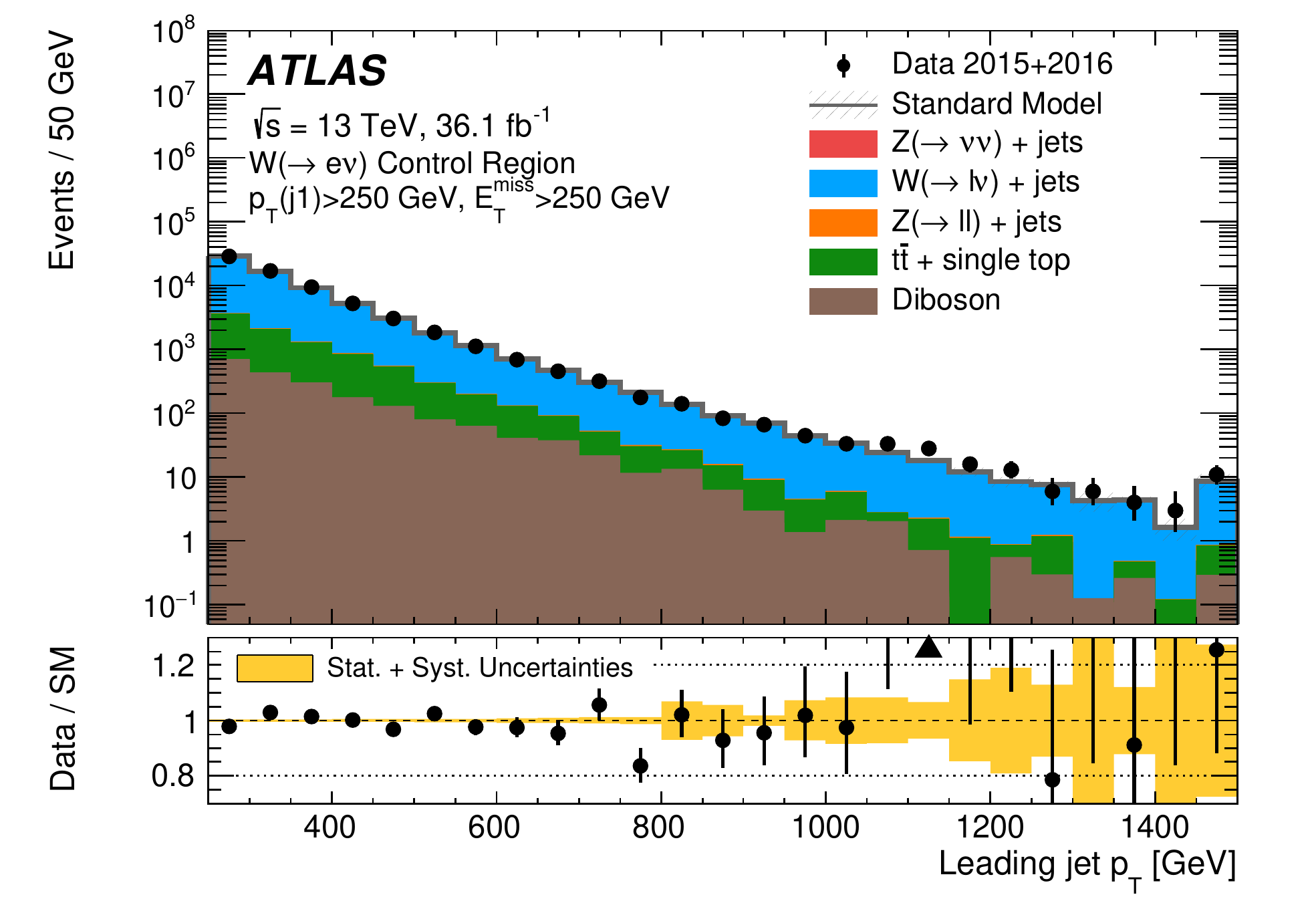}\label{fig2d}
}\\
 \subfloat[][]{
  \includegraphics[width=0.5\textwidth]{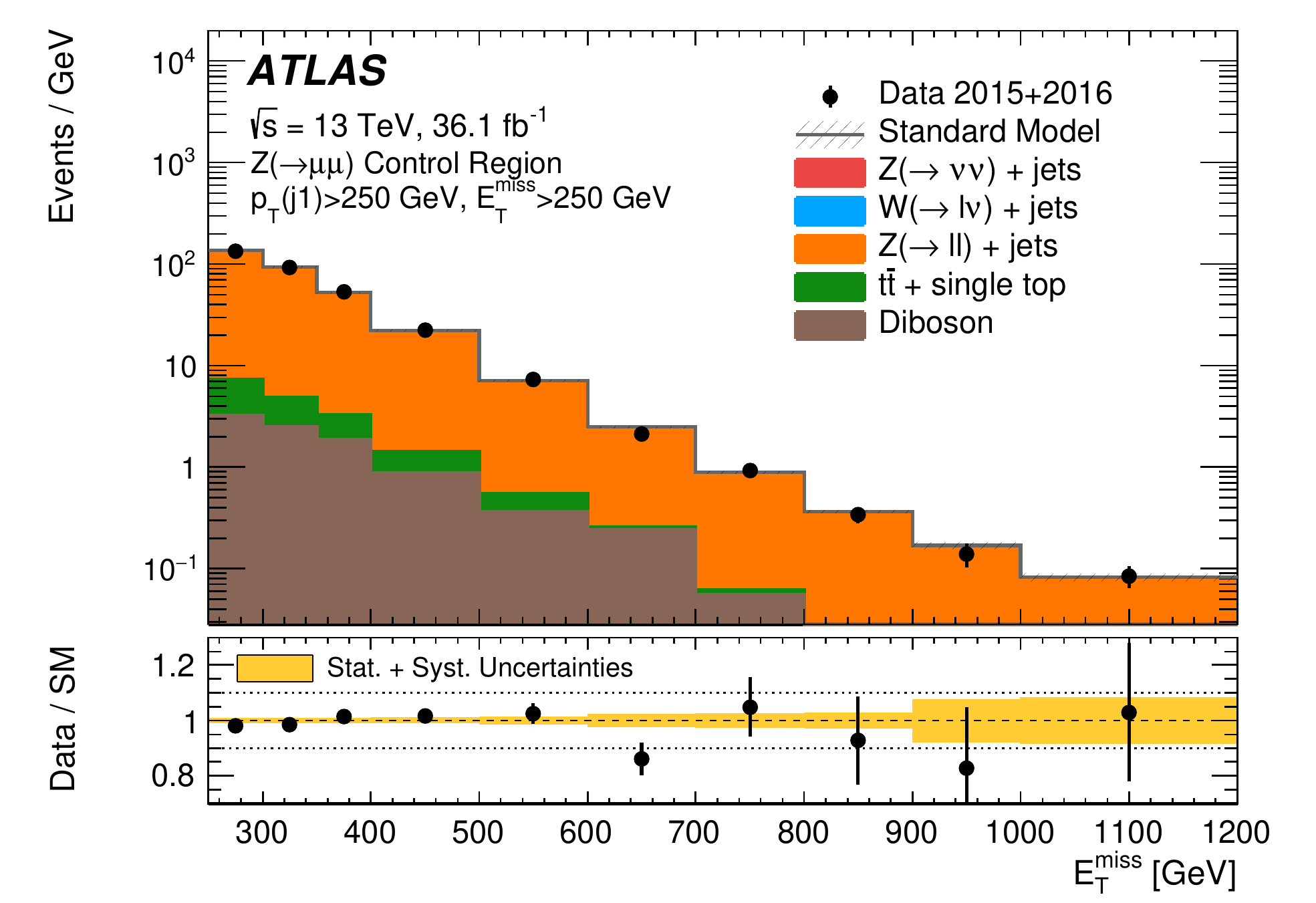}\label{fig2e}
}
 \subfloat[][]{
  \includegraphics[width=0.5\textwidth]{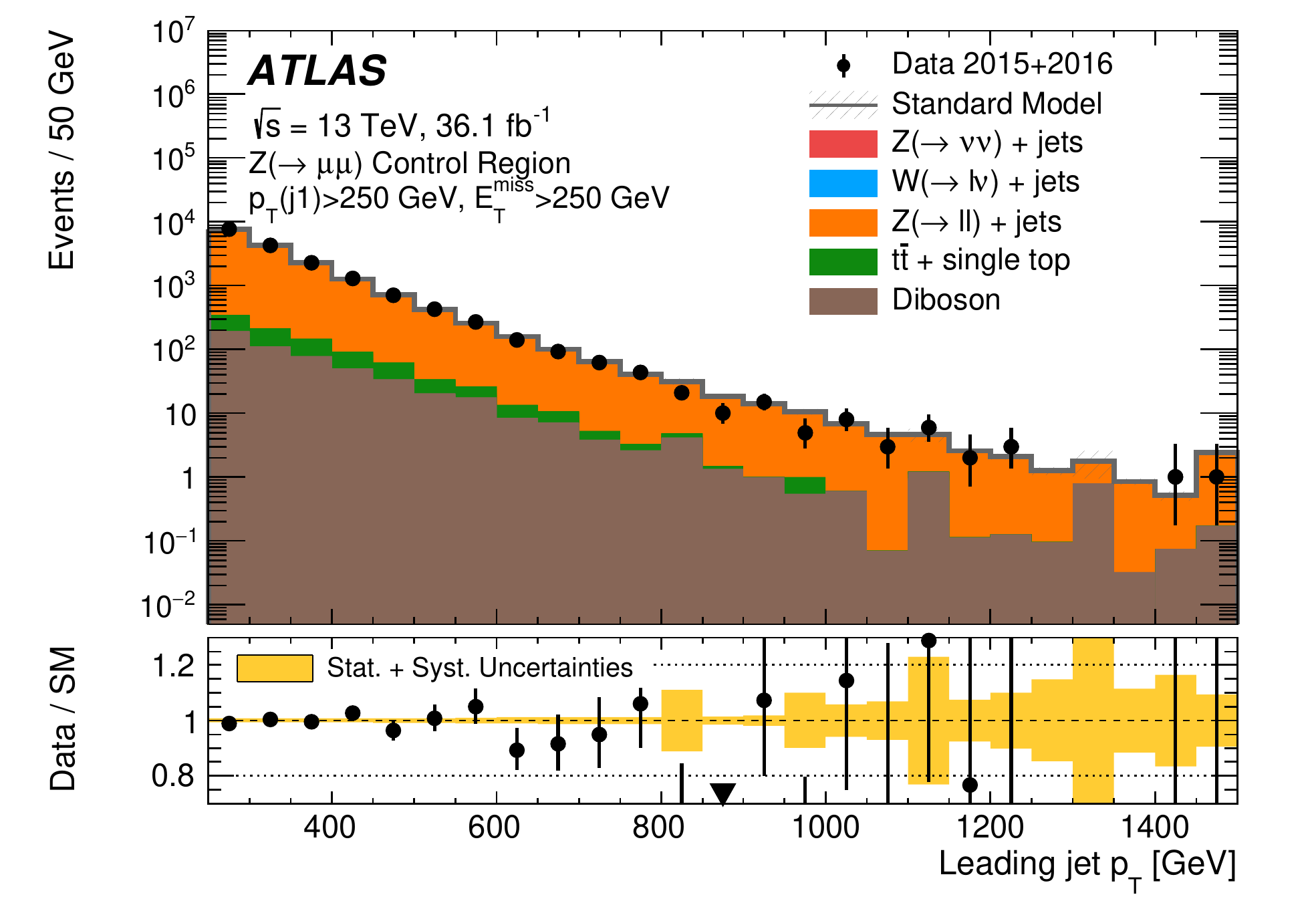}\label{fig2f}
}
\end{center}
\vspace{-.4cm}
\caption{
The measured \protect\subref{fig2a},\protect\subref{fig2c},\protect\subref{fig2e} $\met$  and 
\protect\subref{fig2b},\protect\subref{fig2d},\protect\subref{fig2f} leading-jet $\pt$  distributions
in the $\wmn$+jets,  $\wen$+jets, and $\zmm$+jets control regions, for the $\met > 250~\GeV$ inclusive selection, compared to
the background predictions.
The latter include the global normalization factors extracted from the fit.
The error bands in the ratios include the statistical and systematic  uncertainties in the background predictions as 
determined by the binned-likelihood fit to the data in the control regions.
The last bin of the $\met$ and leading-jet $\pt$ distributions contains overflows.
The contributions from multijet and non-collision backgrounds are negligible and are not shown in the Figures.
}
\label{fig:cr1}
\end{figure}

\FloatBarrier

%
% --- PLOTS CONTROL REGIONS
%

\begin{figure}[!ht]
\begin{center}
 \subfloat[][]{
  \includegraphics[width=0.5\textwidth]{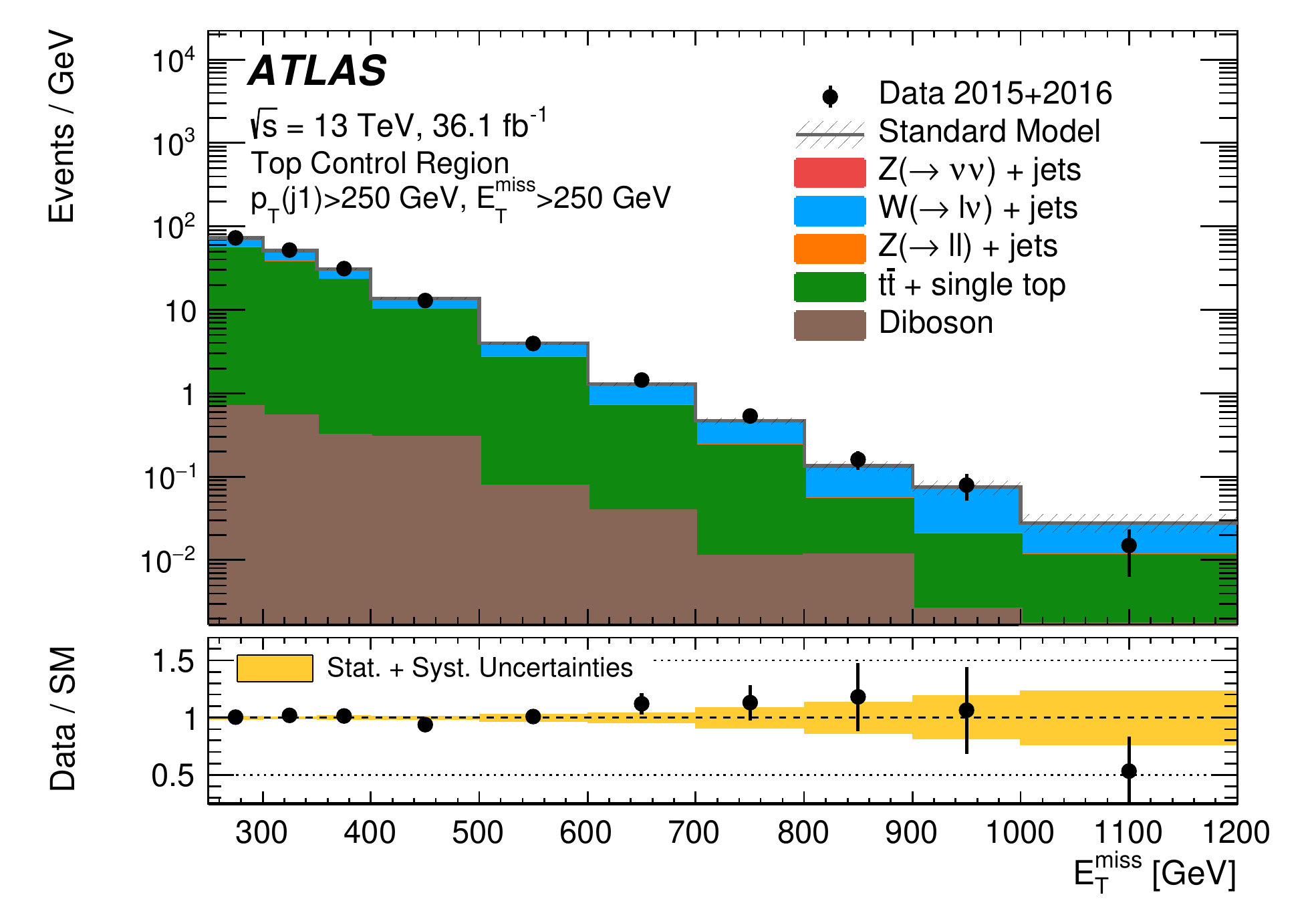}\label{fig3a}
}
 \subfloat[][]{
  \includegraphics[width=0.5\textwidth]{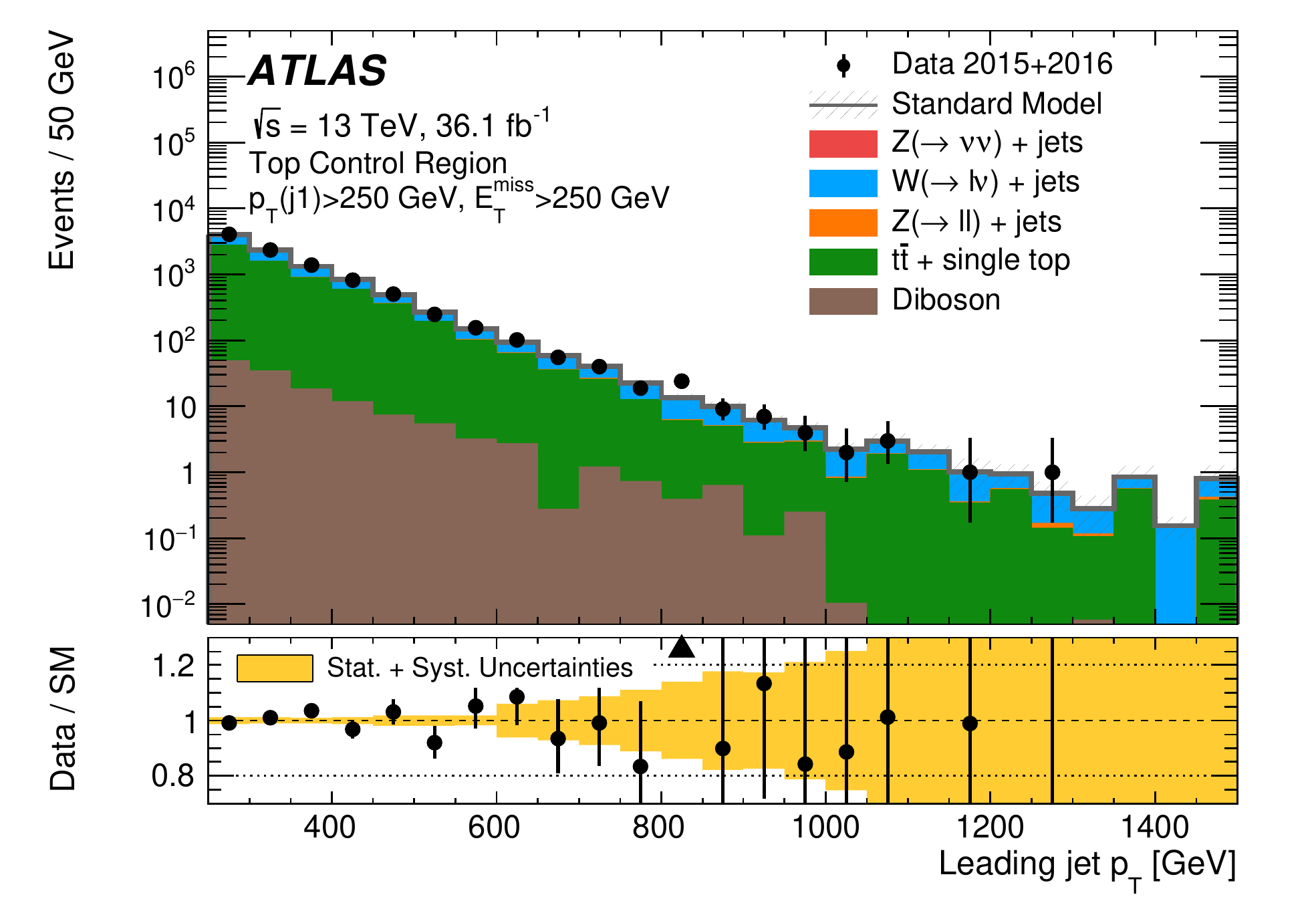}\label{fig3b}
}
\end{center}
\vspace{-.4cm}
\caption{
The measured \protect\subref{fig3a} $\met$ and \protect\subref{fig3b} leading-jet $\pt$  distributions
in the top  control region, for the $\met >250~\GeV$ inclusive selection, compared to
the background predictions.
The latter include the global normalization factors extracted from the fit.
The error bands in the ratios include the statistical and systematic  uncertainties in the background predictions as 
determined by the binned-likelihood fit to the data in the control regions.
The last bin of the $\met$ and leading-jet $\pt$ distributions contains overflows.
The contributions from multijet and non-collision backgrounds are negligible and are not shown in the Figures.
}
\label{fig:cr2}
\end{figure}

% ====================
% SYSTEMATIC UNCERTAINTIES
% ====================
 \section{Systematic uncertainties}
 \label{sec:syst}
 
In this Section, the systematic uncertainties for both the 
background and signal models are presented.  The impacts of the various sources of 
systematic uncertainty on the total background predictions are determined 
by the likelihood fits described in Section~\ref{sec:fitback}.  
Inclusive and exclusive $\met$ selections are considered separately.
For the latter, correlations of systematic uncertainties
across $\met$ bins are taken into account.
The impact of the different sources of uncertainty in representative inclusive $\met$ bins, 
as determined using one-bin likelihood fits, is presented below.  
Experimental and theoretical uncertainties in the signal model predictions are also  presented.

\subsection{Background systematic uncertainties}

Uncertainties in the absolute jet and $\met$ energy scales and 
resolutions~\cite{Aad:2011he} translate into uncertainties in the total 
background which vary between $0.5\%$ for IM1 and $5.3\%$ for 
IM10.
Uncertainties related to jet quality requirements, pile-up 
description and corrections to the jet $\pt$ and $\met$ introduce a 
$0.9\%$ to $1.8\%$ uncertainty in the background predictions.
Uncertainties in the $b$-tagging efficiency, relevant for the definition of the 
$\wmn$+jets and $\ttbar$ control regions, translate into an uncertainty in the total 
background that varies between $0.9\%$ for IM1 and $0.5\%$ for IM10.   
Uncertainties in soft contributions to $\met$ translate into an uncertainty in the total 
background yields that varies between $0.4\%$ for IM1 and $1.7\%$ for IM10.

Uncertainties in
the simulated lepton identification and reconstruction efficiencies, energy/momentum scale and 
resolution~\cite{ATLAS-CONF-2016-024,Aad:2014nim,Aad:2016jkr}  translate 
into an uncertainty in the total background which varies between
$0.2\%$  and $1.7\%$ for IM1 and between $0.3 \%$ and $2.3 \%$ for IM10 selection.

Uncertainties in $W/Z$+jets predictions~\cite{Aaboud:2017hbk,ATL-PHYS-PUB-2017-006}  related to the modelling of parton showers in {\SHERPA} and 
the choice of PDFs translate into an uncertainty in the total background that varies between $0.8\%$
for IM1 and $0.7\%$ for IM10.  Uncertainties on the implementation of higher-order QCD and electroweak 
parton-level calculations in the MC predictions, as described in Ref.~\cite{Lindert:2017olm},  
include: uncertainties in the QCD renomalization/factorization scales, affecting both the normalization and the shape of the 
predicted boson-$\pt$ distribution; uncertainties associated with the non-universality of QCD corrections across $W$+jets and $Z$+jets processes; 
uncertainties in electroweak corrections beyond NNLO, unknown electroweak NLO correction terms at very high boson-$\pt$, and   
limitations of the 
Sudakov approximation adopted in the calculation; uncertainties in the QCD and electroweak interference terms; and uncertainties on the implementation of the higher-order QCD corrections in {\SHERPA}, affected by a limited MC statistics at large boson-$\pt$. Altogether, this translates into an uncertainty in the total background that varies between $0.4\%$ for IM1 and $2\%$ for IM10.  

Theoretical uncertainties in the predicted background yields for 
top-quark-related processes include  
variations in parton-shower parameters and the amount of initial- and final-state 
soft gluon radiation, and the difference between predictions from different MC event generators~\cite{ATL-PHYS-PUB-2017-007}. 
This introduces an 
uncertainty in the total  background of about 
$0.3\%$ for IM1, becoming negligible at very high $\met$.

Uncertainties in the diboson contribution are estimated 
as the difference between the yields of
the {\SHERPA} and {\normalfont \scshape Powheg} event generators~\cite{ATL-PHYS-PUB-2017-005}, after 
taking into account the difference between the cross sections, which is 
then summed in quadrature with a $6\%$ theory uncertainty in the NLO cross section.
This translates into an uncertainty on the total background 
of about $0.2\%$ for IM1 and about $0.8\%$ for IM10.

Uncertainties in the estimation of   multijet and non-collision backgrounds
translate into a  $0.5\%$ uncertainty of the total background for IM1 and
have a negligible impact on the  total background predictions at larger $\met$.
Similarly, the  $3.2\%$ uncertainty in the integrated luminosity is 
included in the fit. It nearly cancels out in the data-driven determination 
of the SM background and translates into an uncertainty in the total background yield 
of about $0.1\%$ for IM1.

% --------------
% --- Signal 
% --------------

\subsection{Signal systematic uncertainties}
\label{sec:signalsys}

Sources of systematic uncertainty in the predicted signal yields are considered separately for each model 
of new physics using a common set of procedures. The procedures are described here, while the numerical uncertainties 
are given with the associated results for each model in Section~\ref{sec:results}.
Experimental uncertainties include those related to the jet and $\met$ reconstruction, 
energy scales and resolutions, and the 
integrated luminosity. 
Other uncertainties related 
to the jet quality requirements are negligible.

Uncertainties affecting the signal acceptance in the generation of 
signal samples include: uncertainties in the  modelling of the initial- and 
final-state gluon radiation, determined using simulated samples with 
modified parton-shower parameters (by factors of two or one half);  
uncertainties due to PDFs and variations of the $\alpha_{\mathrm s}(m_Z)$ 
value employed, as computed from the envelope of CT10, 
MMHT2014~\cite{Harland-Lang:2014zoa} and NNPDF30 error sets; and uncertainties due to the 
choice of renormalization and factorization scales.  In addition, 
theoretical uncertainties in the predicted cross sections, including PDF 
and renormalization- and factorization-scale  uncertainties, are assessed  
separately for the different models.

% ====================
% RESULTS
% ====================
\section{Results and interpretation}
\label{sec:results}
The number of events in the data and the individual background
predictions in several inclusive and exclusive signal regions, as 
determined using 
the background estimation procedure 
discussed in Section~\ref{sec:fitback},  are 
presented in Tables~\ref{tab:srincl} and ~\ref{tab:srexcl}. The results for all the signal regions are summarized in Table~\ref{tab:srsum}.
Good agreement is observed between the data and the SM predictions in each case.
The  SM predictions for the inclusive selections are 
determined with a total uncertainty of $2.4\%$, $2.7\%$,  and $9.7\%$ for the IM1, IM5, and IM10 
signal regions, respectively, which include correlations between uncertainties in the individual 
background contributions.  

Figure~\ref{fig:sr1}  shows
several measured distributions  
compared to the SM predictions in the region $\met > 250~\GeV$, for which 
the normalization factors applied to the MC predictions, and the related uncertainties,  
are determined  from the global fit carried out in exclusive $\met$ bins. 
For illustration  purposes, the distributions  include the impact of 
example
ADD, SUSY, and WIMP scenarios.
In general, the SM predictions provide a good description of the measured distributions. 
The differences observed in the jet multiplicity distribution do not have an impact in the results. 
Statistical tests using the binned profile likelihood fit described above, and considering different scenarios for new physics, give
$p$-values for a background-only hypothesis in the range 0.01--0.04, corresponding to 
agreement with the SM predictions 
within approximately 2.1$\sigma$ to 1.7$\sigma$.

\begin{figure}[!ht]
\begin{center}
\subfloat[][]{
\includegraphics[width=0.5\textwidth]{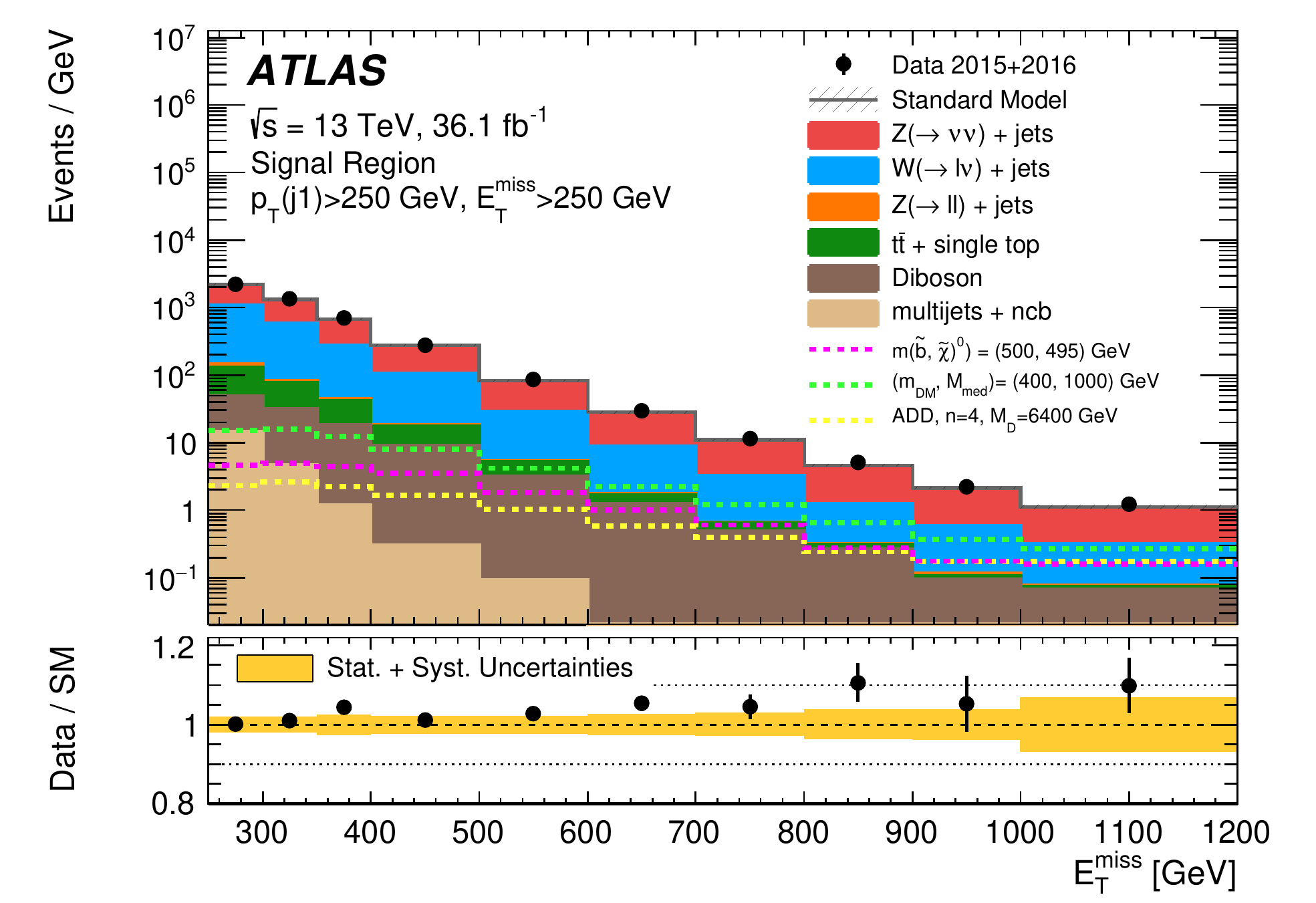}\label{fig4a}
}
\subfloat[][]{
\includegraphics[width=0.5\textwidth]{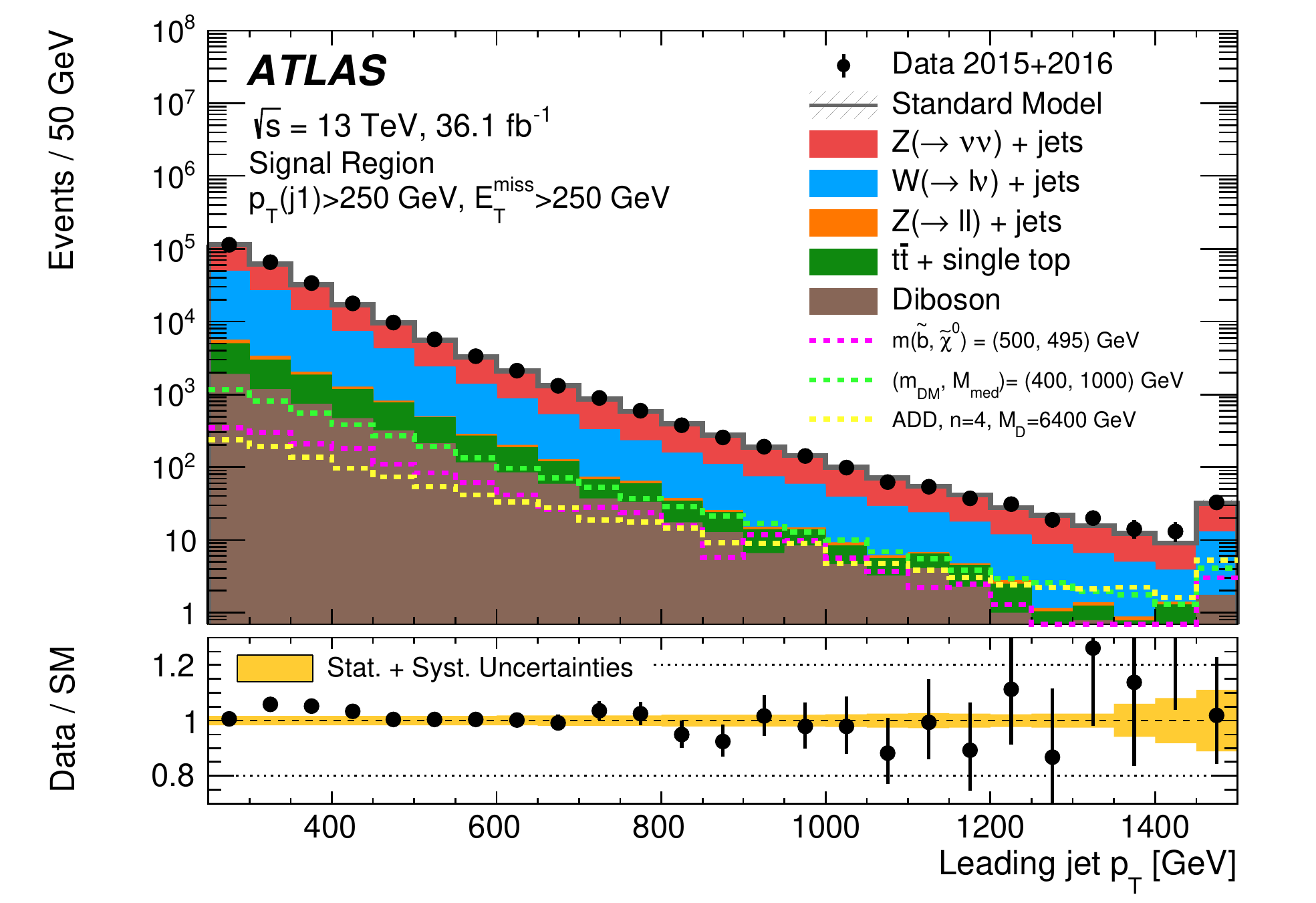}\label{fig4b}
}\\
\subfloat[][]{
\includegraphics[width=0.5\textwidth]{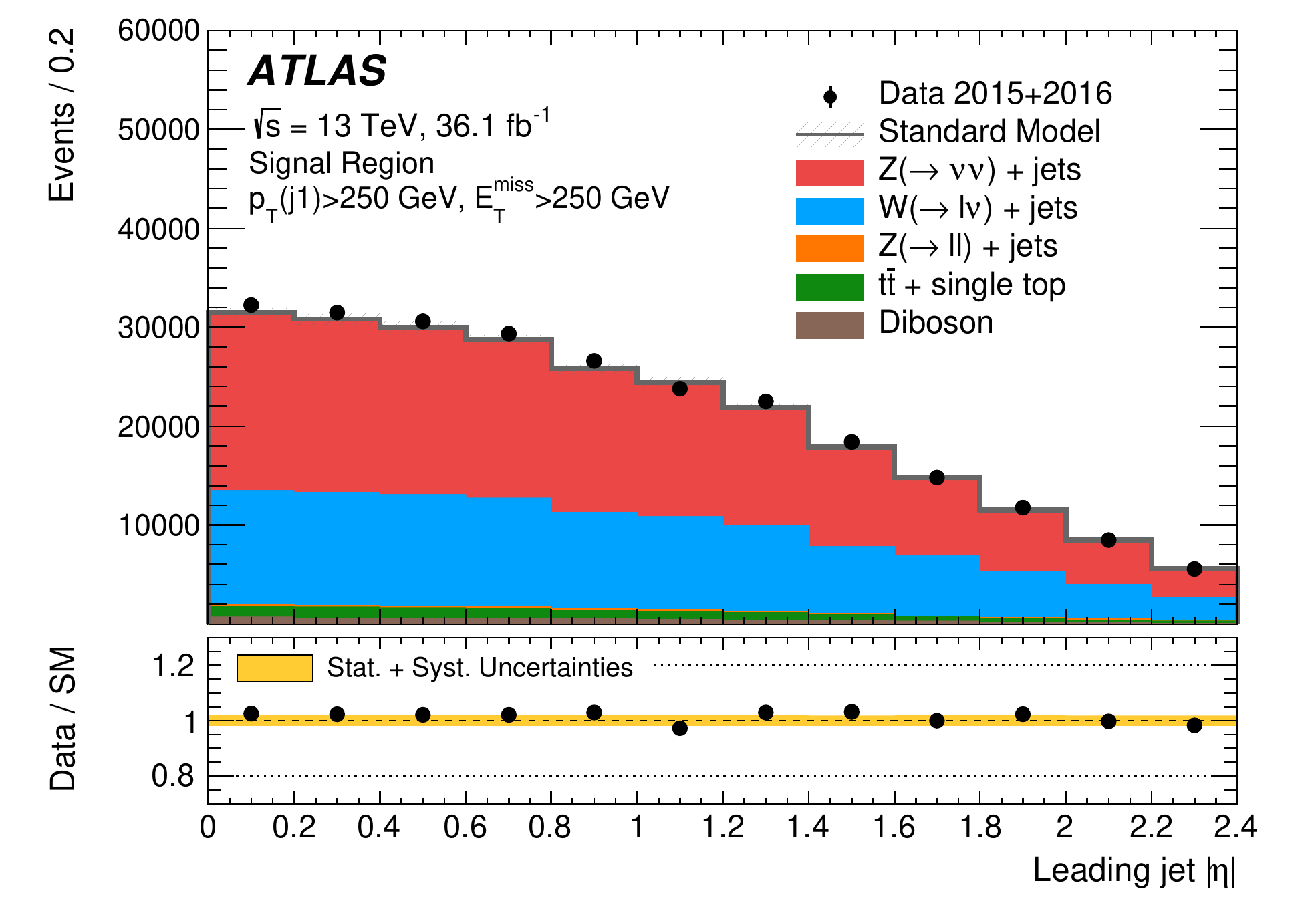}\label{fig4c}
}
\subfloat[][]{
\includegraphics[width=0.5\textwidth]{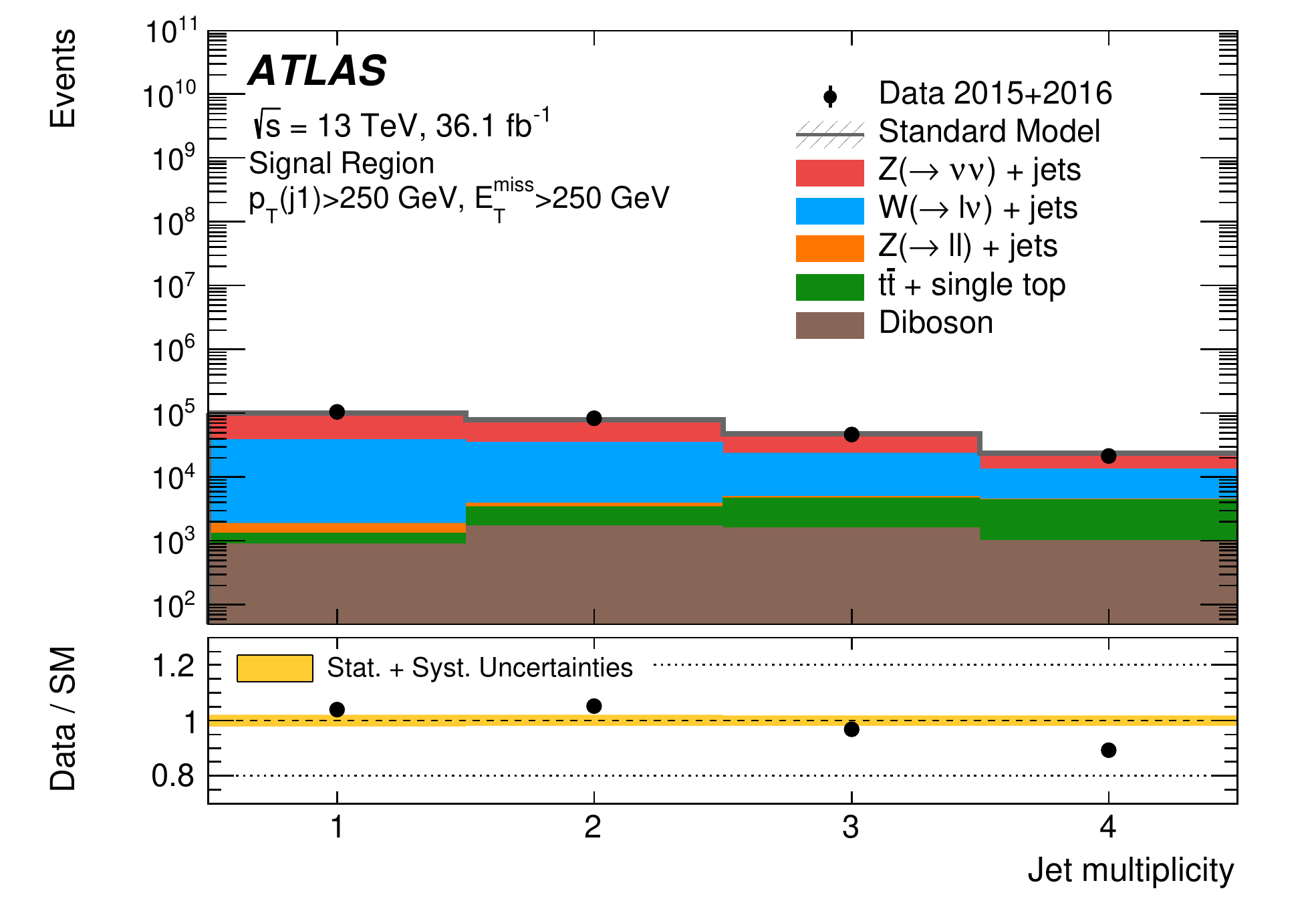}\label{fig4d}
}
\end{center}
\vspace{-.4cm}
\caption{
Measured distributions of the \protect\subref{fig4a} $\met$, \protect\subref{fig4b} leading-jet $\pt$, \protect\subref{fig4c} leading-jet $|\eta|$, and 
\protect\subref{fig4d}
jet multiplicity   
for the $\met >250~\GeV$ selection compared to the SM predictions.  The latter are normalized  with normalization 
factors as determined by the  global fit that considers exclusive $\met$ regions. 
For illustration purposes, the distributions of 
example
ADD, SUSY, and WIMP scenarios  are included.
The error bands in the ratios shown in the lower panels include both the statistical and systematic uncertainties
in the background predictions. 
The last bin of the $\met$ and leading-jet $\pt$ distributions contains overflows.
The contributions from multijet and non-collision backgrounds are negligible and are only shown  in the case of the $\met$ distribution.
}
\label{fig:sr1}
\end{figure}

\begin{table}[!tb]
\caption{Data and SM background predictions in the signal region for several inclusive $\met$ selections, as determined using separate one-bin likelihood fits in the control regions.
For the SM prediction, both the statistical and systematic uncertainties are included.
In each signal region, the individual uncertainties for the different background processes can be correlated,
and do not necessarily add in quadrature to the total background uncertainty.
The dash ``\textendash'' denotes negligible background contributions.
}
\begin{center}
\begin{footnotesize}
\begin{tabularx}{\textwidth}{@{\extracolsep{\fill}}lXr@{\,$\pm$\,}lXr@{\,$\pm$\,}lXr@{\,$\pm$\,}lXr@{\,$\pm$\,}lXr@{\,$\pm$\,}l}\hline
{\normalfont \bfseries  Inclusive Signal Region}& & \multicolumn{2}{c}{~~~IM1} && \multicolumn{2}{c}{~~IM3} && \multicolumn{2}{c}{~~~IM5} && \multicolumn{2}{c}{~~~IM7} && \multicolumn{2}{c}{~IM10}\\
Observed events (36.1 fb${}^{-1}$) && \multicolumn{2}{c}{~~255486}&& \multicolumn{2}{c}{~76808} && \multicolumn{2}{c}{~13680} && \multicolumn{2}{c}{~2122}&& \multicolumn{2}{c}{245}\\ \noalign{\smallskip}\hline\noalign{\smallskip}
SM prediction            && 245900 & 5800  &&  73000 & 1900 &&  12720 & 340  &&  2017 & 90   &&  238 & 23    \\ \cmidrule{3-16}
$\wen$                   &&  20600 & 620   &&   4930 & 220  &&    682 & 33   &&    63 & 8    &&    7 & 2     \\
$\wmn$                   &&  20860 & 840   &&   5380 & 280  &&    750 & 44   &&   115 & 13   &&   17 & 2     \\
$\wtn$                   &&  50300 & 1500  &&  12280 & 520  &&   1880 & 63   &&   261 & 13   &&   24 & 3     \\
$\zee$                   &&   0.11 & 0.03  &&   0.03 & 0.01    &&      \multicolumn{2}{c}{\textendash}    &&     \multicolumn{2}{c}{\textendash}    &&    \multicolumn{2}{c}{\textendash}     \\
$\zmm$                   &&    564 & 32    &&    107 & 9    &&     10 & 1    &&   1.8 & 0.5  &&  0.2 & 0.2   \\
$\ztt$                   &&    812 & 32    &&    178 & 8    &&     24 & 1    &&   3.5 & 0.5  &&  0.4 & 0.1   \\
$\znn$                   && 137800 & 3900  &&  45700 & 1300 &&   8580 & 260  &&  1458 & 76   &&  176 & 18    \\
$\ttbar$, single top     &&   8600 & 1100  &&   2110 & 280  &&    269 & 42   &&    26 & 10   &&    0 & 1     \\
Diboson                  &&   5230 & 400   &&   2220 & 170  &&    507 & 64   &&    88 & 19   &&   13 & 4     \\
Multijet background                &&    700 & 700   &&     51 & 50   &&      8 & 8    &&     1 & 1    &&  0.1 & 0.1   \\
Non-collision background &&    360 & 360   &&     51 & 51   &&      4 & 4    &&     \multicolumn{2}{c}{\textendash}    &&    \multicolumn{2}{c}{\textendash}     \\
\hline
\end{tabularx}
\end{footnotesize}
%%%%%
\end{center}
\label{tab:srincl}
\end{table}
%%%%

%
% ------   
%

\begin{table}[!tb]
\caption{Data and SM background predictions in the signal region for several exclusive $\met$ selections, as determined using a binned likelihood fit in the control regions.
For the SM prediction, both the statistical and systematic uncertainties are included.
In each signal region, the individual uncertainties for the different background processes can be correlated,
and do not necessarily add in quadrature to the total background uncertainty.
The dash ``\textendash'' denotes negligible background contributions.
}
\begin{center}
\begin{footnotesize}
\begin{tabularx}{\textwidth}{@{\extracolsep{\fill}}lXr@{\,$\pm$\,}lXr@{\,$\pm$\,}lXr@{\,$\pm$\,}lXr@{\,$\pm$\,}lXr@{\,$\pm$\,}l}\hline
{\normalfont \bfseries  Exclusive Signal Region} && \multicolumn{2}{c}{~~EM2} && \multicolumn{2}{c}{~~EM4} && \multicolumn{2}{c}{~~~EM6} && \multicolumn{2}{c}{~~EM8} && \multicolumn{2}{c}{~EM9} \\
Observed events  (36.1 fb${}^{-1}$)&& \multicolumn{2}{c}{~67475} && \multicolumn{2}{c}{~27843} && \multicolumn{2}{c}{~2975} && \multicolumn{2}{c}{512} && \multicolumn{2}{c}{223}\\\noalign{\smallskip}\hline\noalign{\smallskip}
SM prediction            && 67100 & 1400&&  27640 & 610       &&  2825 & 78 &&   463 & 19   &&  213 & 9      \\ \cmidrule{3-16}
$\wen$                   &&  5510 & 140 &&  1789 & 59         &&  147 & 9   &&    18 & 1    &&    8 & 1      \\
$\wmn$                   &&  6120 & 200 &&  2021 & 82         &&  173 & 9   &&    21 & 5    &&   11 & 1      \\
$\wtn$                   && 13680 & 310 &&  4900 & 110        &&  397 & 11  &&    55 & 5    &&   29 & 2      \\
$\zee$                   &&  0.03 & 0   && 0.02  & 0.02       &&    \multicolumn{2}{c}{\textendash}   &&     \multicolumn{2}{c}{\textendash}    &&    \multicolumn{2}{c}{\textendash}      \\
$\zmm$                   &&   167 & 8   &&    36 & 2          &&  2.0 & 0.2 &&   0.4 & 0.1  &&  0.5 & 0.1    \\
$\ztt$                   &&   185 & 6   &&    68 & 4          &&  5.1 & 0.3 &&   0.3 & 0.1  && 0.31 & 0.04   \\
$\znn$                   && 37600 & 970 && 17070 & 460        && 1933 & 57  &&   337 & 12   &&  153 & 7      \\
$\ttbar$, single top     &&  2230 & 200 &&   848 & 86         &&   43 & 6   &&     4 & 1    &&  1.3 & 0.4    \\
Diboson                  &&  1327 & 90  &&   874 & 64         &&  124 & 16  &&    26 & 5    &&   10 & 2      \\
Multijet background                &&   170 & 160 &&    13 & 13         &&    1 & 1   &&     1 & 1    &&  0.1 & 0.1    \\
Non-collision background &&    71 & 71  &&    18 & 18         &&    \multicolumn{2}{c}{\textendash}   &&     \multicolumn{2}{c}{\textendash}    &&    \multicolumn{2}{c}{\textendash}      \\
\hline
\end{tabularx}
\end{footnotesize}
%%%%%                                                                                                                                                                          
\end{center}
\label{tab:srexcl}
\end{table}
%%%%                                     

\begin{table}[!tb]
\caption{Data and SM background predictions in the signal region for the different selections.
For the SM predictions both the statistical and systematic uncertainties are included.
}
\begin{center}
\begin{footnotesize}
\begin{tabularx}{\textwidth}{@{\extracolsep{\fill}}lXr@{\,$\pm$\,}lXrX|lXr@{\,$\pm$\,}lXr}\hline
\multicolumn{7}{c}{{{\normalfont \bfseries Inclusive Signal Region}}} & \multicolumn{6}{c}{{{\normalfont \bfseries Exclusive Signal Region}}} \\
Region&&\multicolumn{2}{c}{Predicted}&&Observed &&Region && \multicolumn{2}{c}{Predicted}&&Observed \\ \hline 
IM1  && 245900 & 5800  && 255486&& EM1 && 111100 & 2300 && 111203 \\
IM2  && 138000 & 3400  && 144283&& EM2 &&  67100 & 1400 && 67475  \\
IM3  &&  73000 & 1900  && 76808 && EM3 &&  33820 & 940  && 35285  \\
IM4  &&  39900 & 1000  && 41523 && EM4 &&  27640 & 610  && 27843  \\
IM5  &&  12720 & 340   && 13680 && EM5 &&   8360 & 190  && 8583   \\
IM6  &&   4680 & 160   && 5097  && EM6 &&   2825 & 78   && 2975   \\
IM7  &&   2017 & 90    && 2122  && EM7 &&   1094 & 33   && 1142   \\
IM8  &&    908 & 55    && 980   && EM8 &&    463 & 19   && 512    \\
IM9  &&    464 & 34    && 468   && EM9 &&    213 & 9    &&  223   \\
IM10 &&    238 & 23    && 245   && EM10 &&    226 & 16   &&  245   \\
\hline
\end{tabularx}
\end{footnotesize}
%%%%%
\end{center}
\label{tab:srsum}
\end{table}
%%%%            

The levels of  agreement between the data and the SM predictions for the total number of events
in inclusive and exclusive signal regions are translated into upper limits for the presence of
new phenomena, using a simultaneous likelihood fit in both the control and signal regions,  
and the ${\mathrm CL}_{\mathrm s}$ modified frequentist approach~\cite{Read:2002hq}.   
The inclusive regions are used to set model-independent exclusion limits, and
the exclusive regions are used for the interpretation of the results
within different models of new physics. 
In general, the observed exclusion limits are worse than the expected sensitivity due to  the slight excess of events in 
the data compared to the SM predictions, as shown in Table~\ref{tab:srsum}.

\subsection{Model-independent exclusion limits}

A likelihood fit is performed separately for each 
of the inclusive regions IM1--IM10. As a result,
model-independent observed and expected 95$\%$ confidence level (CL) upper limits 
on the visible cross section, defined as the 
product of production cross section, acceptance and efficiency
$\sigma \times A \times \epsilon$, are extracted from the ratio between 
the 95$\%$ CL upper limit on the number of signal events and the integrated luminosity,     
taking into consideration
the systematic 
uncertainties in the SM backgrounds 
and the  uncertainty in the  integrated luminosity.
The results are presented in Table~\ref{tab:indep}. 
Values of $\sigma \times A \times \epsilon$  
above 531~fb (for IM1) and above 1.6~fb (for IM10)  are excluded at 95$\%$ CL.  

\begin{table}[!ht]
\caption{
Observed and expected 95$\%$ CL upper limits on the  
number of signal events, $S_{\mathrm obs}^{95}$ and  $S_{\mathrm exp}^{95}$, and on the                                                              
visible cross section, defined as the product of cross section, acceptance and efficiency,  
$\langle{\mathrm \sigma}\rangle_{\mathrm obs}^{95}$,  
for the IM1--IM10 selections.}             
\begin{center}
\setlength{\tabcolsep}{0.0pc}
{\footnotesize
\begin{tabularx}{\textwidth}{@{\extracolsep{\fill}}lXrXrXrl}
\noalign{\smallskip}\hline\noalign{\smallskip}
{\normalfont \bfseries Selection} && $\langle{\mathrm \sigma}\rangle_{\mathrm obs}^{95}$~[fb]  &&  $S_{\mathrm obs}^{95}$  && \multicolumn{2}{c}{$S_{\mathrm exp}^{95}$} \\ 
\noalign{\smallskip}\hline\noalign{\smallskip}
IM1   &&  $531$ &&  19135  && 11700 & ${}^{+4400}_{-3300}$ \\  \noalign{\smallskip}
IM2   &&  $330$ &&  11903  &&  7000 & ${}^{+2600}_{-2600}$ \\  \noalign{\smallskip}
IM3   &&  $188$ &&  6771   &&  4000 & ${}^{+1400}_{-1100}$ \\  \noalign{\smallskip}
IM4   &&  $93$  &&  3344   &&  2100 & ${}^{+770}_{-590}$   \\  \noalign{\smallskip}
IM5   &&  $43$  &&  1546   &&   770 & ${}^{+280}_{-220}$   \\  \noalign{\smallskip}
IM6   &&  $19$  &&  696    &&   360 & ${}^{+130}_{-100}$   \\  \noalign{\smallskip}
IM7   &&  $7.7$ &&  276    &&   204 & ${}^{+74}_{-57}$     \\  \noalign{\smallskip}
IM8   &&  $4.9$ &&  178    &&   126 & ${}^{+47}_{-35}$     \\  \noalign{\smallskip}
IM9   &&  $2.2$ &&  79     &&    76 & ${}^{+29}_{-21}$     \\  \noalign{\smallskip}
IM10  &&  $1.6$ &&  59     &&    56 & ${}^{+21}_{-16}$     \\  \noalign{\smallskip}\hline
\end{tabularx}
}
%%%%
\end{center}
\label{tab:indep}
  \end{table}

%
% -- WIMPS
%

\subsection{Weakly interacting massive particles}
The results are translated into exclusion limits on WIMP-pair production. Different simplified models are  
considered  with the exchange of an 
axial-vector, vector  or a pseudoscalar mediator 
in the $s$-channel.  In addition, a model with the exchange of a coloured scalar mediator is 
considered, as described in Section~\ref{sec:intro}.      

In the case of the exchange of an axial-vector mediator, and for 
WIMP-pair production with $m_{Z_A}>2m_\chi$, typical 
$A \times \epsilon$ values for the signal models  with a $1~\TeV$ mediator range 
from 25\%  to 0.4\% for IM1 and IM10 selections, respectively. Very similar 
values are obtained in the case of the vector mediator, whereas $A \times \epsilon$ values 
in the range between $32 \%$ and $1 \%$ are computed for the pseudoscalar mediator model 
with $m_{Z_P} = 1~\TeV$ and $m_\chi = 10~\GeV$.
Finally, in the case of the coloured scalar 
mediator, $A \times \epsilon$ values in the range from $35 \%$ to $0.7 \%$ are obtained 
for IM1 and IM10 selections, respectively, for a mediator mass of 
$1~\TeV$ and 
$m_\eta \gg m_\chi$.

The experimental uncertainties related to the jet and $\met$ scales and 
resolutions introduce similar uncertainties in the signal yields for axial-vector, vector and pseudoscalar models.  
They vary between $2\%$ and $7\%$ for the IM1 selection  and between $3\%$ and $9\%$ for the IM10 selection, depending on the parameters of the model. 
In the case of the coloured scalar mediator model, these uncertainties vary between
2$\%$ and 6$\%$ for IM1 and between 4$\%$ and about 10$\%$ for IM10. 
The uncertainty related to the modelling of the initial- and final-state radiation 
translates into a $20\%$ uncertainty in the signal acceptance, common to all the $s$-channel models.  
In the case of the coloured scalar mediator model, this uncertainty varies between 10$\%$ and 30$\%$, depending on the 
kinematic selection. 
The choice of different PDF sets results in up to a $20\%$ 
uncertainty in the acceptance and up to a $10\%$ uncertainty in the cross section, depending on the model considered. 
Varying the renormalization and factorization scales introduces up to $25\%$ 
variations of the cross section and up to  $10\%$ change in the acceptance, depending on the model considered. In 
addition,  the uncertainty in the integrated luminosity is included.

A simultaneous fit to the signal and control regions in the exclusive $\met$ bins 
is performed, and used to set observed and expected 95\% CL exclusion limits on the 
parameters of the model.
Uncertainties in the signal acceptance times 
efficiency, the background predictions, and the luminosity are considered, and 
correlations between systematic uncertainties in signal and background 
predictions are taken into account. The fit accounts for the contamination of 
the control regions by signal events which a priori is estimated to be very small.

Figure~\ref{fig:dm2}\protect\subref{fig5a}  shows the observed and expected 95\% CL exclusion 
contours in the $m_{Z_A}$--$m_\chi$
parameter plane for a simplified model with an 
axial-vector mediator, Dirac WIMPs, and couplings $g_q = 1/4$ and $g_\chi = 1$. 
In addition, observed limits are shown using 
$\pm1\sigma$ theoretical  uncertainties in the signal cross sections. In the 
on-shell regime, the models with mediator masses up to $1.55~\TeV$ are excluded for 
$m_\chi=1~\GeV$. For $m_\chi < 1~\GeV$, the monojet analysis maintains its sensitivity for excluding DM models.
This analysis loses sensitivity to the models in the off-shell regime, where 
cross sections are suppressed due to the virtual production of the mediator.
Perturbative unitarity is 
violated in the parameter region defined by $m_\chi > \sqrt{\pi/2}\ m_{Z_A}$ 
\cite{Kahlhoefer:2015bea}. The masses corresponding to the relic density~\cite{Boveia:2016mrp}  
as determined by the Planck and WMAP 
satellites~\cite{Adam:2015rua,2013ApJS..208...19H}, within the WIMP dark-matter model and in the absence of any 
interaction other than the one considered, are indicated in the Figure as a line 
that crosses the excluded region at $m_{Z_A} \sim 1200~\GeV$ and $m_\chi \sim 440~\GeV$. 
The region towards lower WIMP masses or higher mediator masses corresponds to 
dark-matter overproduction. 

\begin{figure}[htb]
\centering
  \subfloat[][]{
\includegraphics[width=0.495\textwidth]{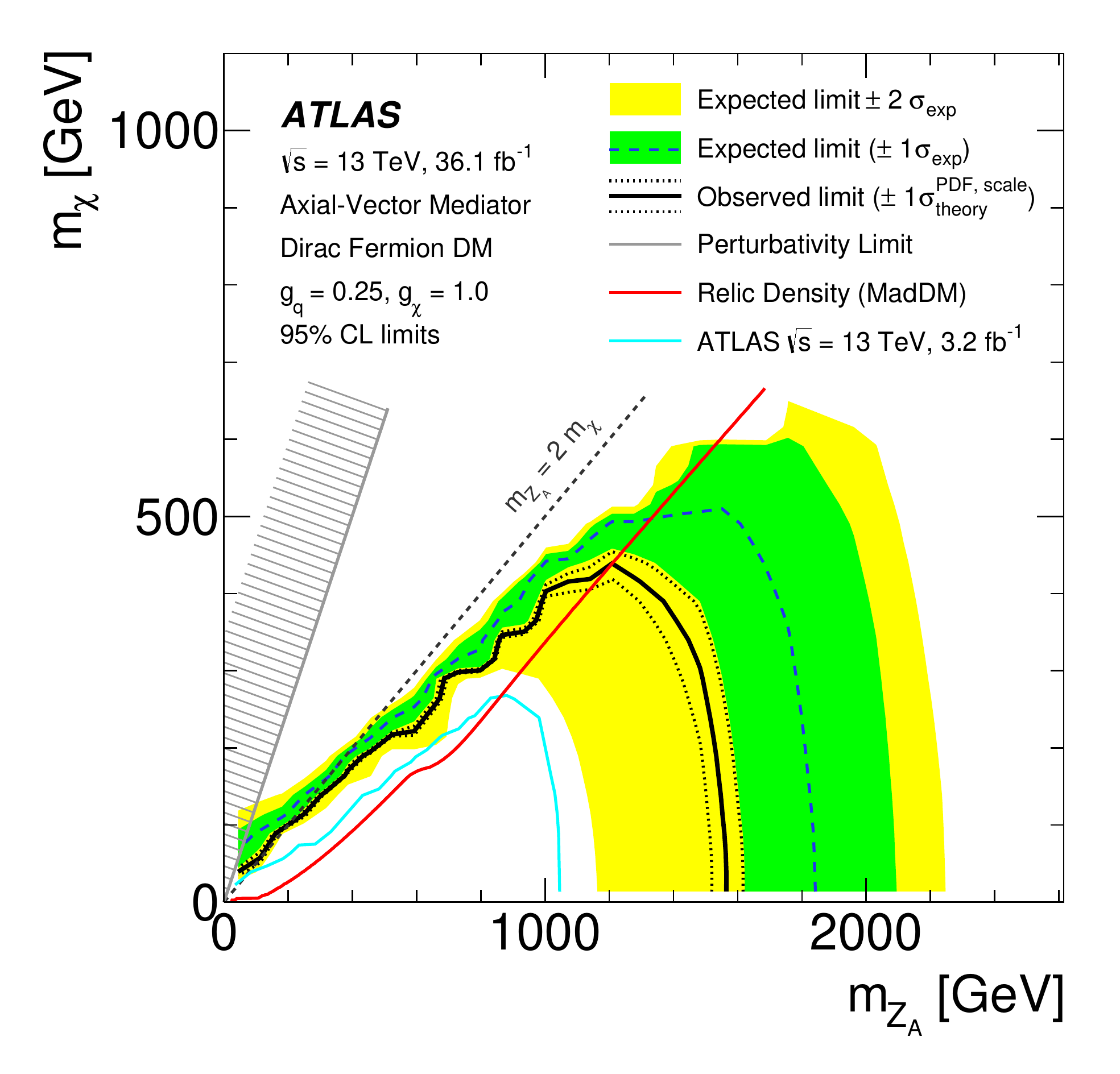}\label{fig5a}
}
  \subfloat[][]{
\includegraphics[width=0.495\textwidth]{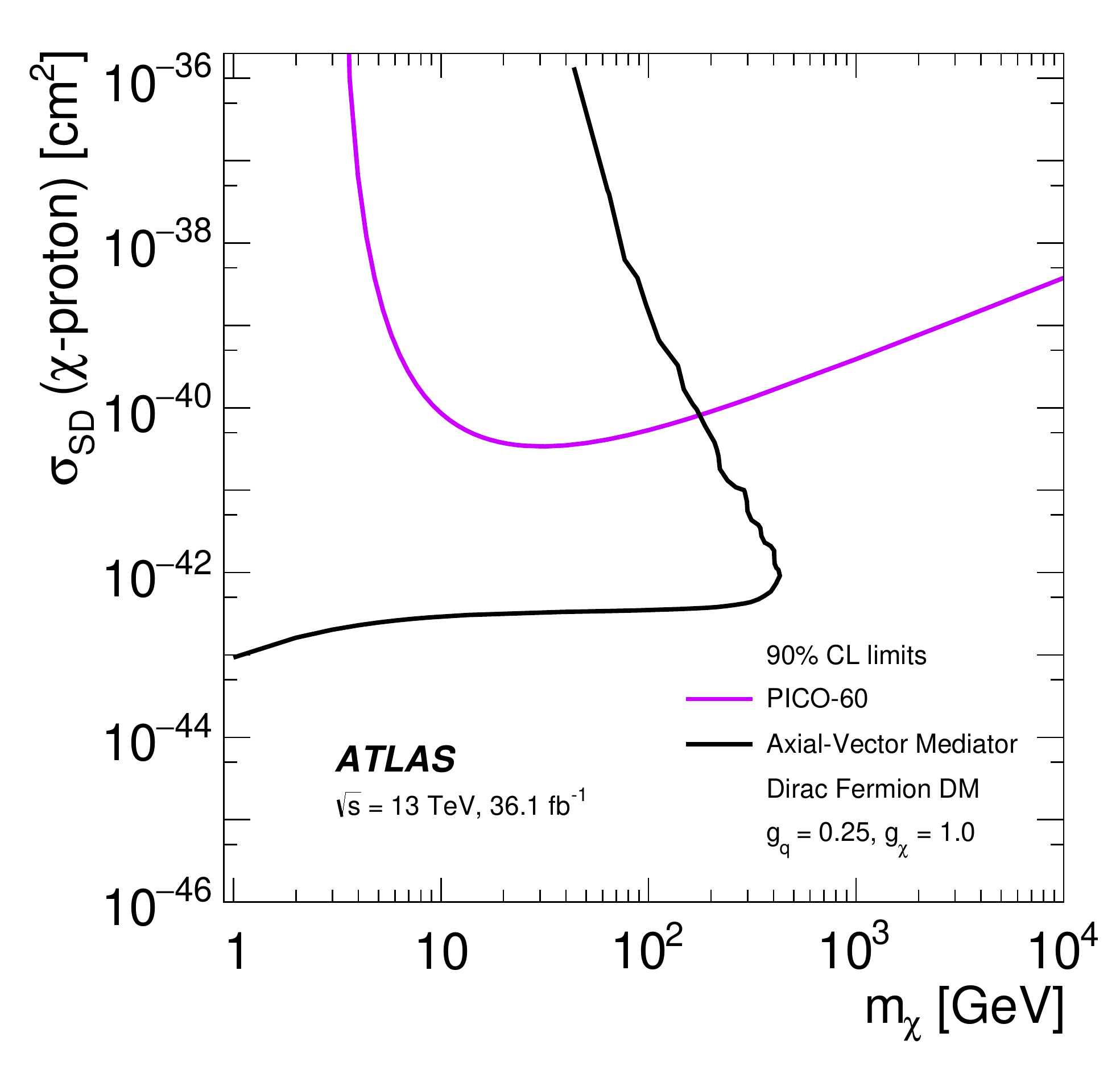}\label{fig5b}
}
\caption{
\protect\subref{fig5a} 
Axial-vector 95\% CL exclusion contours in the 
$m_{Z_A}$--$m_\chi$
parameter plane. The 
solid (dashed) curve shows the 
observed (expected) limit, while the 
bands indicate the $\pm1\sigma$  theory uncertainties in the observed limit and $\pm1\sigma$ and 
$\pm2\sigma$ ranges of the expected limit in the absence of a signal. 
The red curve corresponds to the set of points for which the expected relic density is consistent with the WMAP measurements (i.e. $\Omega h^2 = 0.12$), as computed with {\normalfont \scshape MadDM}~\cite{Backovic:2015cra}. The region on the right of the curve corresponds to higher 
predicted relic abundance than these measurements.
The region excluded due to 
perturbativity, defined by $m_\chi > \sqrt{\pi/2}\ m_{Z_A}$, is indicated by the 
hatched area. The dotted line indicates the kinematic limit for on-shell production $m_{Z_A} = 2 \times m_\chi$. 
The cyan line indicates previous results at $13~\TeV$~\cite{Aaboud:2016tnv} using 3.2~$\ifb$. 
\protect\subref{fig5b}  A comparison of the inferred limits (black line) to the constraints from 
direct detection experiments (purple line) on the spin-dependent WIMP--proton scattering cross 
section in the context of the simplified model with axial-vector 
couplings. Unlike in the 
$m_{Z_A}$--$m_\chi$
parameter plane, the limits are shown at 
90\% CL. The results from this analysis, excluding the region to the left of the 
contour, are compared with limits from 
the 
PICO \cite{Amole:2017dex} experiment.  
The comparison is model-dependent and solely valid in the context of this model, 
assuming minimal mediator width and the coupling values $g_q=1/4$ and $g_\chi=1$.
}
\label{fig:dm2}
\end{figure}

The results are translated into 90\% CL exclusion 
limits on the spin-dependent WIMP--proton scattering  cross section $\sigma_{\mathrm{SD}}$ as a function 
of the WIMP mass, following the prescriptions 
from
Refs.~\cite{Buchmueller:2014yoa,Boveia:2016mrp}. Among results from different 
direct-detection experiments, in Figure~\ref{fig:dm2}\protect\subref{fig5b}  the 
exclusion limits obtained in this analysis
are compared to the most stringent limits  from 
the PICO direct-detection experiment~\cite{Amole:2017dex}.
The limit at the maximum value of the WIMP–-proton scattering cross section displayed corresponds to the lowest excluded values 
$m_{Z_A} = 45~\GeV$ and $m_\chi = 45~\GeV$  
of the mediator and dark matter masses displayed in Figure~\ref{fig:dm2}\protect\subref{fig5a}.
This 
comparison is model-dependent and solely valid in the context of this particular 
model. In this case, stringent limits on the scattering cross section 
of the order of $2.9 \times 10^{-43}$~cm${}^2$ ($3.5 \times 10^{-43}$~cm${}^2$)
for WIMP masses below $10~\GeV$ ($100~\GeV$) are 
inferred from this analysis, and complement the results from direct-detection 
experiments for $m_\chi < 10~\GeV$.  The kinematic loss of model sensitivity 
is expressed by the turn of the WIMP exclusion line, reaching 
back to low WIMP masses and intercepting the exclusion lines from the 
direct-detection experiments at around $m_\chi=200~\GeV$.

In Figure~\ref{fig:dmV_2d}, the results are translated into 95$\%$~CL exclusion contours in the 
$m_{Z_V}$--$m_\chi$
parameter plane for the simplified model with a 
vector mediator, Dirac WIMPs, and couplings $g_q = 1/4$ and $g_\chi = 1$.
The results are obtained from those for the axial-vector model, taking into account the cross-section differences between models, motivated 
by the fact that 
the two models present compatible particle-level selection acceptances.
For very light WIMPs, mediator masses below about $1.55~\TeV$ are excluded. As in the case of the axial-vector mediator model, 
in the regime $m_{Z_V} < 2 m_\chi$, the sensitivity for exclusion
is drastically reduced to low mass differences below $400~\GeV$ in $m_\chi$. 

\begin{figure}[htb]
 \centering
 \includegraphics[width=9cm]{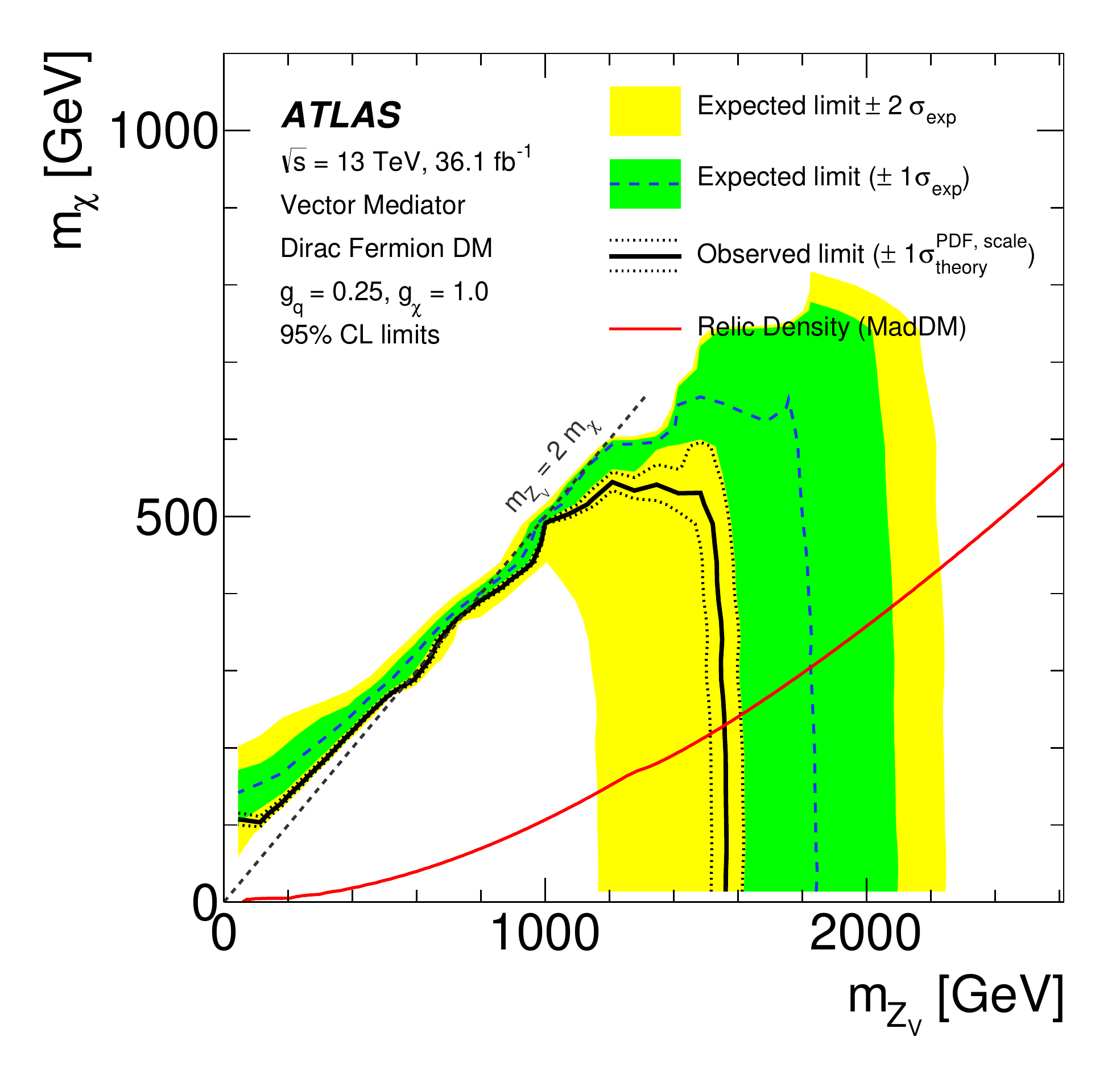}
 \caption{Observed (solid line) and expected (dashed line) exclusions at $95\%$ CL on the vector mediator models with 
$g_{q}=1/4,g_{\chi}=1.0$ and minimal mediator width, as a function of the assumed mediator and DM masses. The regions within the drawn contours are excluded. 
The red curve corresponds to the set of points for which the expected relic density is consistent with the WMAP measurements 
(i.e. $\Omega h^2 = 0.12$), as computed with {\normalfont \scshape MadDM}~\cite{Backovic:2015cra}. The region on the right of the curve corresponds to higher
predicted relic abundance than these measurements.
The dotted line indicates the kinematic limit for on-shell production $m_{Z_V} = 2 \times m_\chi$. 
}
 \label{fig:dmV_2d}
\end{figure}

\begin{figure}[htb]                                                                                                                                                     
\centering                                                                                                                                               \subfloat[][]{            
\includegraphics[width=0.495\textwidth]{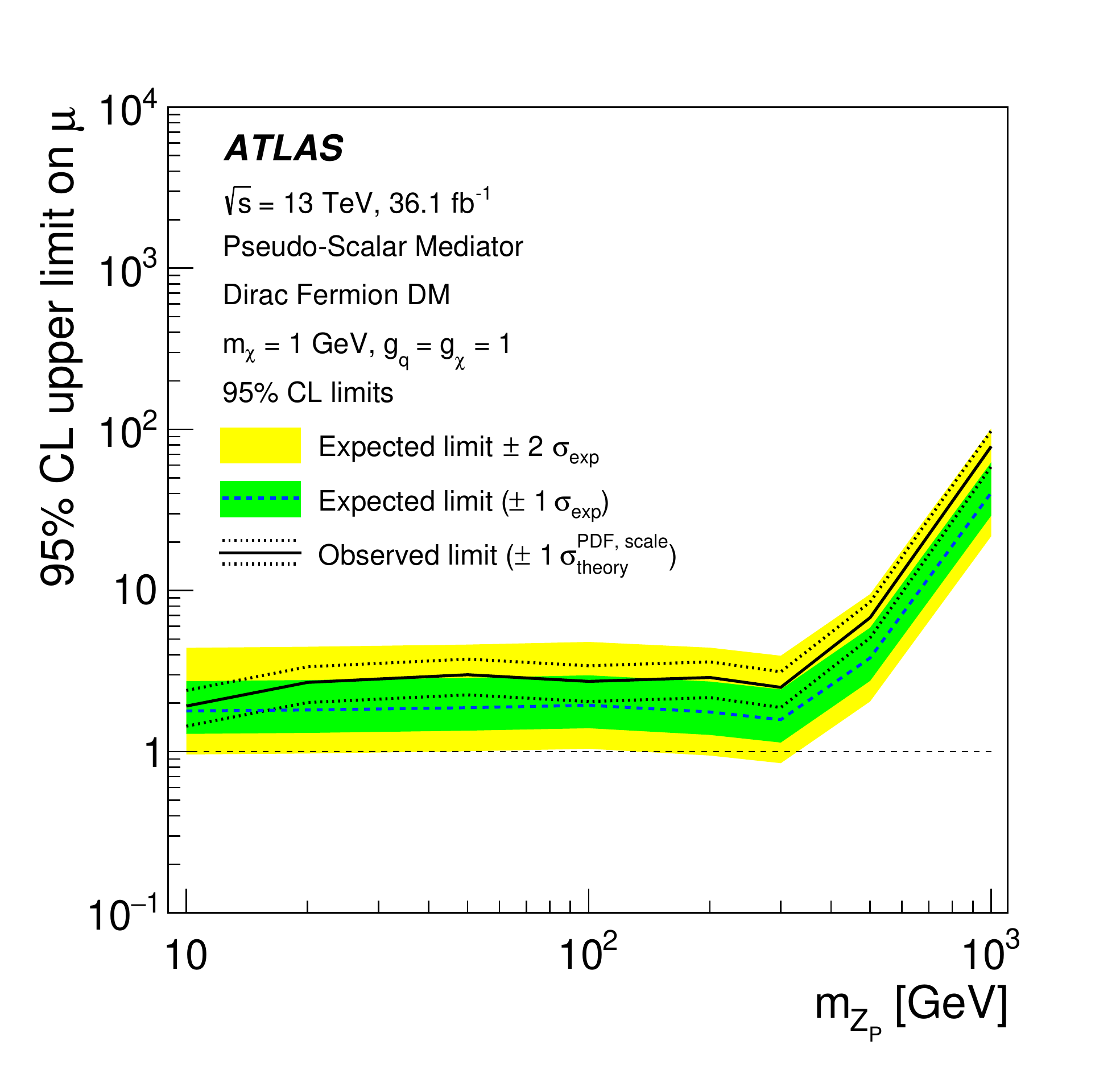}\label{fig7a}
  }
  \subfloat[][]{            
\includegraphics[width=0.495\textwidth]{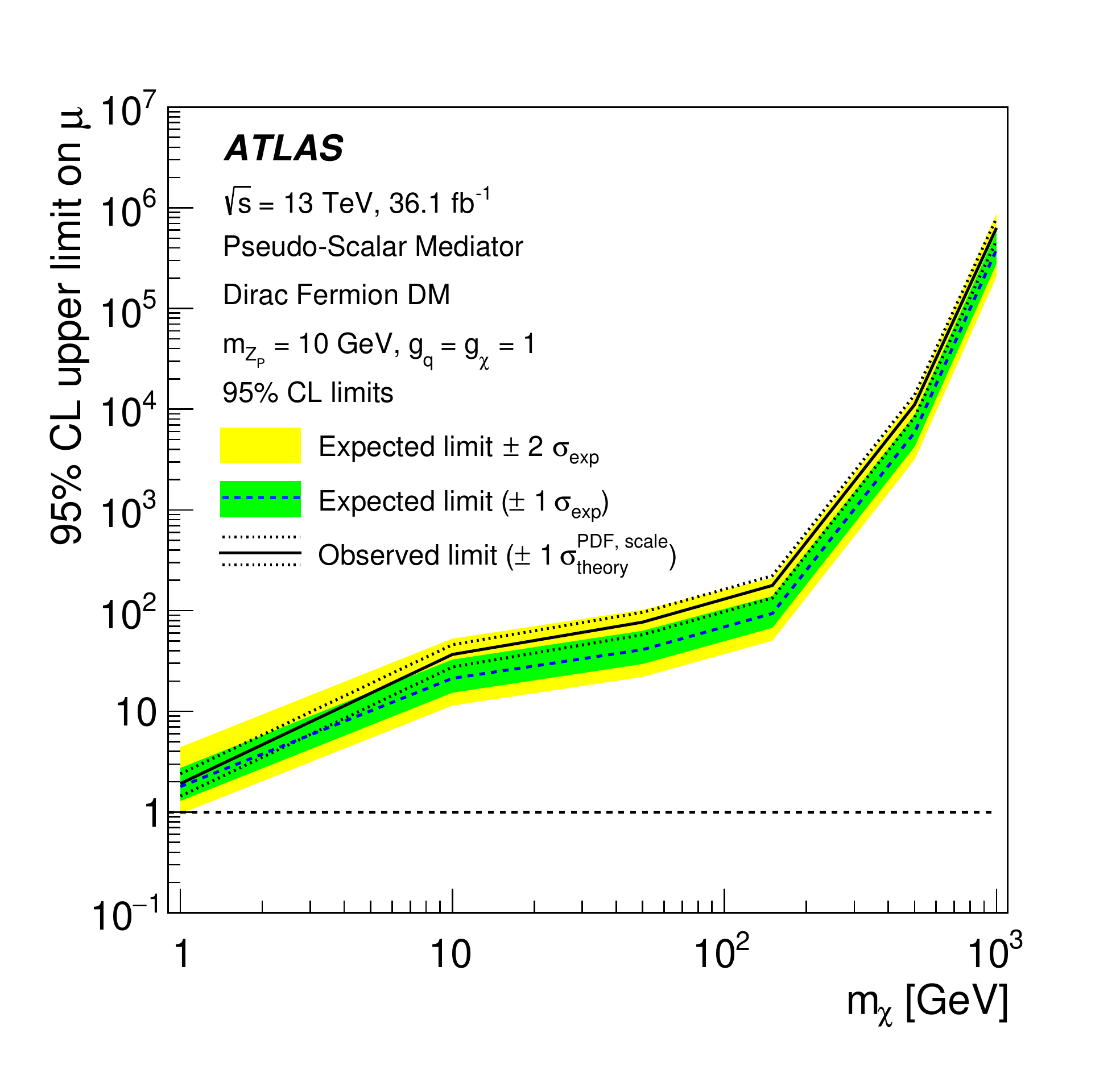}\label{fig7b}                                                                            }
\caption{Observed and expected 95 $\%$ CL limits on the signal strength $\mu \equiv \sigma^{95\% \ \mathrm{CL}}/\sigma$ as 
a function of \protect\subref{fig7a} the mediator mass for a very light WIMP and \protect\subref{fig7b} 
the WIMP mass for  $m_{Z_P}=10~\GeV$, in a 
model with spin-0 pseudoscalar mediator and $g_{q}=g_{\chi}=1.0$. 
The bands indicate the $\pm 1\sigma$  theory uncertainties in the
observed limit and the $\pm 1\sigma$  and $\pm 2\sigma$ ranges of the expected limit in the absence of a signal.
}                                                                                          
\label{fig:running}          
\end{figure}

The simplified model with a pseudoscalar mediator was considered with 
couplings to quarks and dark matter equal to unity. 
For WIMP masses in the range 0--300~$\GeV$ and $m_{Z_P}$ in the range 
0--700~$\GeV$, the analysis does not yet have enough sensitivity.
As an example, Figure~\ref{fig:running} presents the analysis 
sensitivity in terms of 95$\%$ CL limits on the signal strength, 
$\mu \equiv \sigma^{95\% \ \mathrm{CL}}/\sigma$,   
as a function of $m_{Z_P}$, for very light WIMPs, and as a function 
of $m_\chi$, for $m_{Z_P}= 10~\GeV$. For mediator masses below 
$300~\GeV$ and very light WIMPs, cross sections of the order of 
2-to-3 times larger than that of the  corresponding signal are 
excluded. For mediator masses above $300~\GeV$ or larger dark-matter 
masses, the sensitivity of the analysis to this particular model vanishes rapidly.  

Finally, Figure~\ref{fig:tdm} presents the observed and expected 95\% CL 
exclusion contours in the 
$m_\eta$--$m_\chi$
parameter plane for the dark-matter production model with a coloured 
scalar mediator, Dirac WIMPs, and couplings set to $g = 1$.  Mediator 
masses up to about $1.67~\TeV$  are excluded at 95$\%$ CL for light 
dark-matter particles. In the case of  $m_\chi =  m_\eta$, masses up 
to $620~\GeV$ are excluded.                                                                                                                                   
\begin{figure}[htb]
\centering   
\includegraphics[width=9cm]{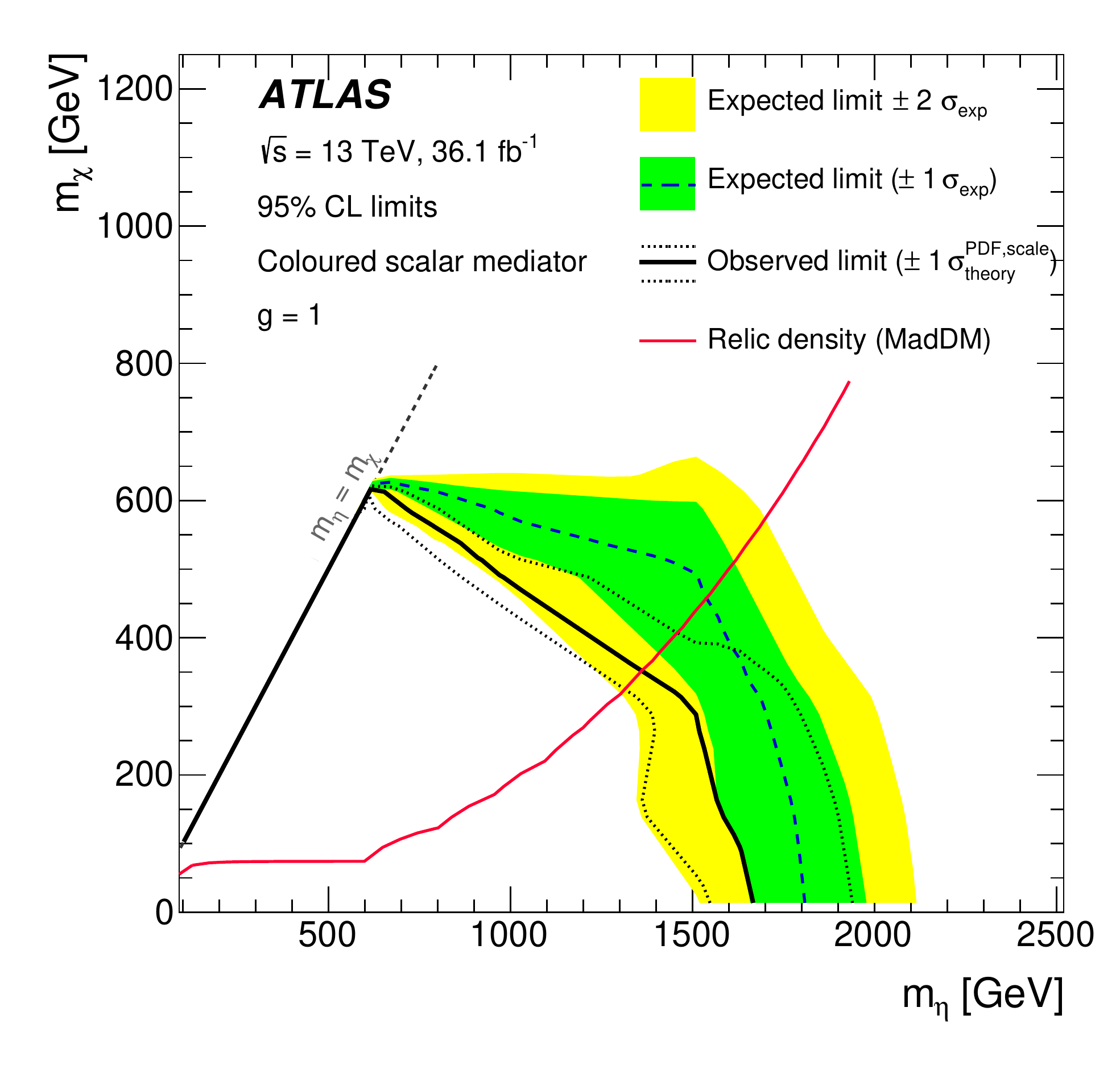}       
\caption{
Exclusion contours at 95$\%$ CL in the 
$m_\eta$--$m_\chi$
parameter plane for the coloured scalar mediator model, with minimal width 
and  coupling set to $g = 1$. The solid (dashed) curve shows the 
observed (expected) limit, while the bands indicate the 
$\pm1\sigma$  theory uncertainties in the observed limit and 
$\pm1\sigma$ and $\pm2\sigma$ ranges of the expected limit in the 
absence of a signal. The red curve corresponds to the expected relic 
density, as computed with {\normalfont \scshape MadDM}~\cite{Backovic:2015cra}. 
The kinematic limit for the mediator on-shell production 
$m_\eta =  m_\chi$, is indicated by the dotted line. }       
\label{fig:tdm} 
\end{figure}

%
% -- SUSY
%

\subsection{Squark-pair production}

Different models of squark-pair production are considered: stop-pair production 
with $\tilde{t}_1 \to  c +  \tilde{\chi}^{0}_{1}$, stop-pair production with 
$\tilde{t}_1 \to  b + ff^{'} + \tilde{\chi}^{0}_{1}$, sbottom-pair production
with $\tilde{b}_1 \to b + \ninoone$, and squark-pair production with $\tilde{q} 
\to q + \ninoone$ ($q =u,d,c,s$). In each case separately, the results are 
translated into exclusion limits as a function of the squark mass for different 
neutralino masses.

The results are translated into exclusion limits on the pair production cross section of top 
squarks with $\tilde{t}_1 \to  c +  \tilde{\chi}^{0}_{1}$ (with branching fraction B=100$\%$) as a 
function of the stop mass for different neutralino masses.  The typical value of  
$A \times \epsilon$ of the selection criteria varies, with increasing stop and 
neutralino masses, between 0.7$\%$ and 1.4$\%$ for IM1,  and between $0.04\%$ 
and 1.3$\%$ for IM10.
Observed and  expected 95$\%$ CL exclusion limits are set as in the case of the WIMP models.  In addition, observed limits are 
computed using $\pm 1\sigma$ variations of the theoretical predictions for the 
SUSY cross sections.        

The uncertainties related 
to the jet and $\met$ scales and resolutions introduce uncertainties in the 
signal yields which vary between $1\%$ and $3\%$ for different 
selections and  squark and neutralino masses.  In addition, the uncertainty in 
the integrated luminosity is included. The uncertainties related to the modelling 
of initial- and final-state gluon radiation translate into a $7\%$ to 
$17\%$ uncertainty in the signal yields. The uncertainties due to the PDFs 
result in a $5\%$ to $17\%$ uncertainty in the signal yields.  Finally, 
the variations of the renormalization and factorization scales introduce a 
$4\%$ to $13\%$ uncertainty in the signal yields.

Figure~\ref{fig:stop}\protect\subref{fig9a}  presents the results in the case of the $\tilde{t}_1 \to  
c +  \tilde{\chi}^{0}_{1}$ decays. The previous limits from the ATLAS 
Collaboration~\cite{Aaboud:2016tnv}, corresponding to a luminosity of 3.2~$\ifb$, are also shown.  
This analysis has significantly higher sensitivity at
very low 
stop--neutralino mass difference. In the compressed scenario with the stop and neutralino nearly 
degenerate in mass, the exclusion extends up to stop masses of $430~\GeV$.  The 
region with stop--neutralino mass differences below $5~\GeV$ is not considered in the exclusion since in this 
regime the stop could become long-lived. 
\begin{figure}[h]
\begin{center}
  \subfloat[][]{
  \includegraphics[width=0.5\textwidth]{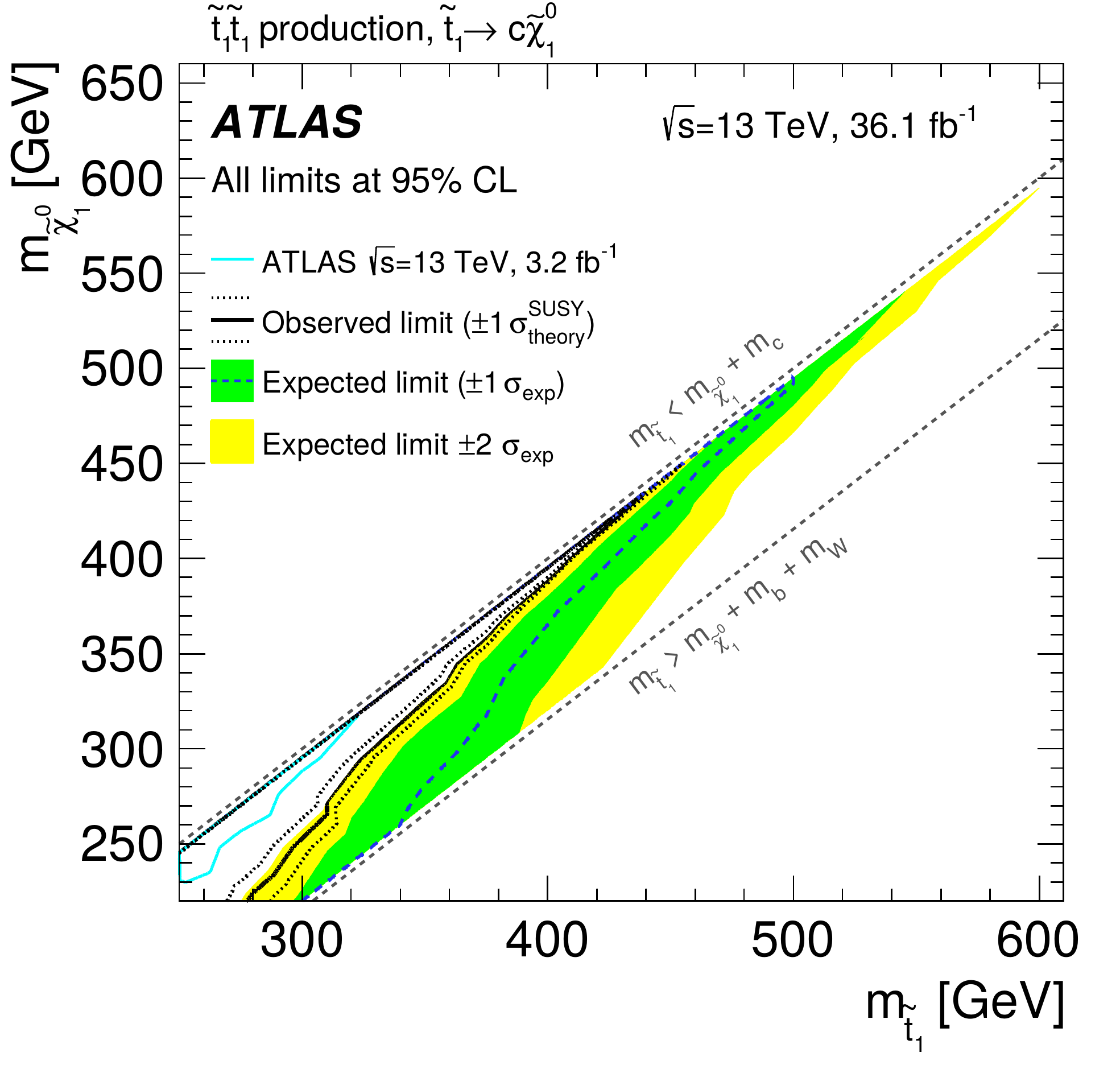}\label{fig9a}
}
  \subfloat[][]{
  \includegraphics[width=0.5\textwidth]{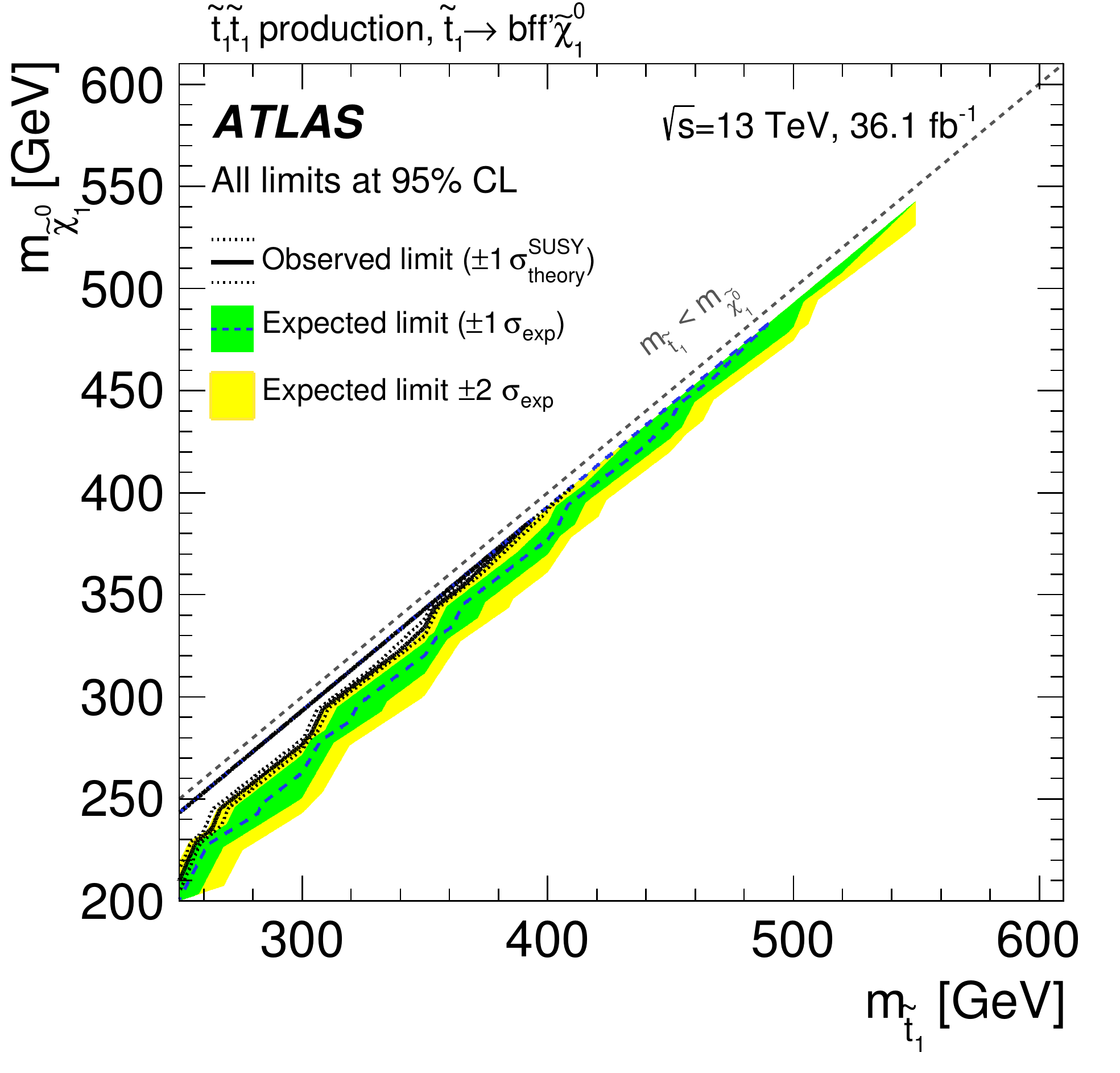}\label{fig9b}
}
\end{center}
\caption{
Excluded regions at the 95$\%$ CL in   the ($\tilde{t}_1,\tilde{\chi}^{0}_{1}$) mass 
plane for \protect\subref{fig9a} the decay channel $\tilde{t}_1 \to  c +  \tilde{\chi}^{0}_{1}$ (B = 
100$\%$) and \protect\subref{fig9b}  the decay channel 
 $\tilde{t}_1 \to  b + ff^{'} + \tilde{\chi}^{0}_{1}$ (B=100$\%$).
The dotted lines around the observed limits indicate the range of observed limits                                     
corresponding to $\pm 1\sigma$ variations of the NLO SUSY cross-section                                              
predictions.
The bands around the expected limits indicate 
the expected $\pm1\sigma$ and $\pm 2\sigma$ ranges of limits in the absence of a signal. 
The 
results from this analysis are compared to previous results from the ATLAS 
Collaboration at $\sqrt{s} = 13~\TeV$~\cite{Aaboud:2016tnv} using $3.2 \ \ifb$.
}
\label{fig:stop}
\end{figure}
Figure~\ref{fig:stop}\protect\subref{fig9b}  shows the observed and expected 95$\%$ CL exclusion limits 
as a function of the stop and neutralino masses for the  $\tilde{t}_1 \to b + 
ff^{'} + \tilde{\chi}^{0}_{1}$ (B=100$\%$) decay channel. For  $m_{\tilde{t}_1} - m_{\ninoone}  \sim m_b$, stop 
masses up to $390~\GeV$ are excluded at 95$\%$ CL.  

Figure~\ref{fig:sbottom}\protect\subref{fig10a}  presents the  observed and expected 95$\%$ CL exclusion 
limits as a function of the sbottom and neutralino masses for the  $\tilde{b}_1 
\to b + \tilde{\chi}^{0}_{1}$ (B=100$\%$) decay channel. In the 
scenario with $m_{\tilde{b}_1} - m_{\ninoone} \sim m_b$, this analysis extends 
the 95$\%$ CL exclusion limits up to a sbottom mass of $430~\GeV$. In the case 
of light neutralinos with $m_{\ninoone} \sim 1~\GeV$, sbottom masses up to 
$610~\GeV$ are excluded at 95$\%$ CL. 
\begin{figure}[h]
\begin{center}
  \subfloat[][]{
\includegraphics[width=0.5\textwidth]{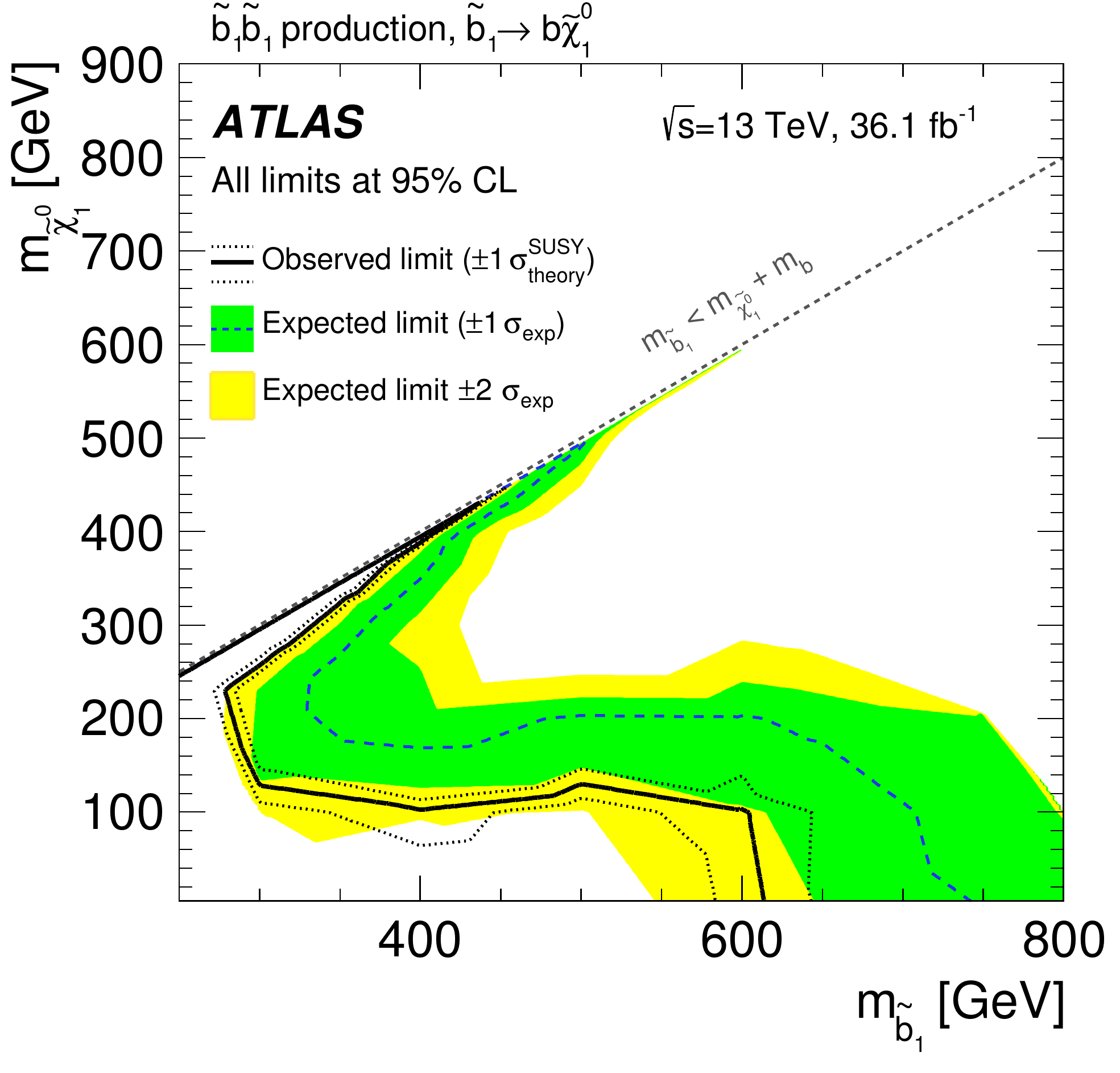}\label{fig10a}
}
  \subfloat[][]{
\includegraphics[width=0.5\textwidth]{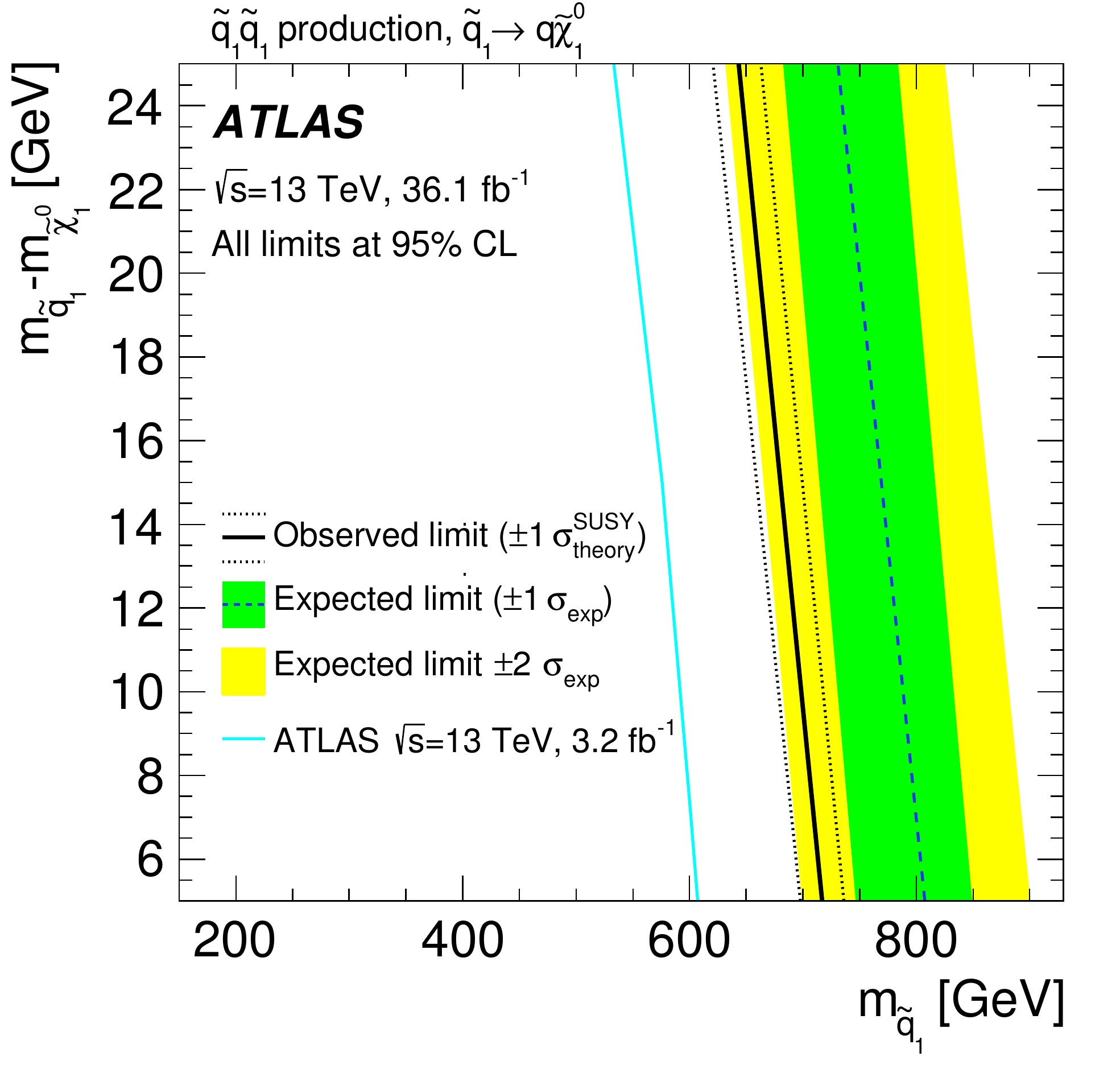}\label{fig10b}
}
\end{center}
\caption{
\protect\subref{fig10a}  Exclusion plane at 95$\%$ CL as a function of sbottom and neutralino masses for 
the decay channel  $\tilde{b}_1 \to  b +  \tilde{\chi}^{0}_{1}$ (B=100$\%$). 
\protect\subref{fig10b}  Exclusion region at 95$\%$ CL as a function of squark mass and the
squark--neutralino mass difference for $\tilde{q} \to q + \tilde{\chi}^{0}_{1}$
($q =u,d,c,s$).
The dotted lines around the observed limit indicate the range of observed limits
corresponding to $\pm 1\sigma$ variations of the NLO SUSY cross-section
predictions.
The bands around the expected limit indicates the expected $\pm1\sigma$ and $\pm 2\sigma$ ranges 
of limits in the absence of a signal.  
The
results from this analysis are compared to previous results from the ATLAS
Collaboration at $\sqrt{s} = 13~\TeV$~\cite{Aaboud:2016tnv} using $3.2 \ \ifb$.
} 
\label{fig:sbottom}
\end{figure}
Finally, Figure~\ref{fig:sbottom}\protect\subref{fig9b}   presents the observed and expected 95$\%$ CL 
exclusion limits as a function of the squark mass and the squark--neutralino mass 
difference for $\tilde{q} \to q + \ninoone$ ($q =u,d,c,s$). In the compressed 
scenario with similar squark and neutralino masses, squark masses below $710~\GeV$ 
are excluded at 95$\%$ CL.
These results
are a significant improvement on previous exclusion limits~\cite{Aaboud:2016tnv}, and 
complement inclusive SUSY searches~\cite{Aaboud:2016zdn} in  such  mass-compressed regime.

%
% --- ADD LED
%

\subsection{Large extra spatial dimensions}

The level of agreement between the data and the SM predictions is also translated into limits  on the parameters of 
the ADD model, as described in Section~\ref{sec:intro}.
Only the signal regions with $\met > 400~\GeV$, 
where the SM background is moderate and the shape difference between signal and the SM background becomes apparent, 
have 
sufficient sensitivity to ADD signal.
The typical value of $A \times \epsilon$ of the selection criteria varies, as the number of extra dimensions $n$ increases from $n=2$ to $n=6$, 
between 13$\%$ and 17$\%$ for IM4  and between $0.8\%$ and 1.4$\%$ for IM10.

The effect of experimental uncertainties related to jet and $\met$
scales and resolutions is found to be similar to the effect in the WIMP models.
The uncertainties related to the modelling of the initial- and final-state 
gluon radiation  
translate into 
uncertainties in the ADD signal acceptance which  
vary between $11\%$ and $13\%$ with increasing $\met$ and approximately independent of $n$.
The  uncertainties due to the
PDFs, affecting the predicted signal yields, increase from $11\%$ at $n=2$ to $43\%$ at $n=6$.
Similarly, the variations of the renormalization and factorization scales 
introduce a $23\%$ to $36\%$ uncertainty 
in the signal yields, with increasing $n$. 

Observed and expected 95$\%$ CL exclusion limits are set as in the case of the WIMP and SUSY models.
The $- 1\sigma$ variations of the ADD
theoretical cross sections result in about a 7$\%$ to 10$\%$ decrease in the nominal observed limits, depending on $n$.
Figure~\ref{fig:add} and Table~\ref{tab:add} present the results.
Values of $M_D$ below $7.7~\TeV$ at $n=2$ and below $4.8~\TeV$ at $n=6$ are excluded at
95$\%$~CL,  
which improve on the exclusion limits from previous results using 3.2~$\ifb$ of $13~\TeV$ data~\cite{Aaboud:2016tnv}.

As discussed in 
Refs.~\cite{ATLAS:2012ky, Aad:2015zva}, 
the  analysis partially probes the phase-space region with $\hat{s} > M_D^2$, 
where $\sqrt{\hat{s}}$ is the
centre-of-mass energy of the hard interaction. This challenges the validity of the model implementation and the lower bounds
on $M_D$, as they depend on  the unknown ultraviolet behaviour of the effective theory.  
The observed 95$\%$ CL limits are recomputed after 
suppressing, with a weighting  factor $M_D^4/\hat{s}^2$, the signal events with $\hat{s} > M_D^2$, here 
referred to as damping.  This results in a negligible decrease of the quoted 95$\%$  CL lower limits on $M_D$, 
as also shown in Table~\ref{tab:add_limits}.

\begin{table}[!hb]
\caption{
The $95\%$ CL observed and expected lower limits on the fundamental Planck 
scale in $4+n$ dimensions, $M_D$, as a function of the number of extra 
dimensions $n$, considering nominal LO signal cross sections.
The impact of the $\pm 1\sigma$  theoretical uncertainty on
the observed limits and the  expected $\pm 1\sigma$ range of limits in the 
absence of a signal are also given. Finally, the $95\%$ CL observed limits 
after damping of the signal cross section for  $\hat{s} > M_D^2$ (see text) 
are quoted.}
  \label{tab:add_limits}
\begin{center}
\begin{footnotesize}
  \begin{tabular*}{\textwidth}{@{\extracolsep{\fill}}lccc} \hline\noalign{\smallskip}
    \multicolumn{4}{c}{{\normalfont \bfseries ADD Model Limits on $M_{D}$
    } (95\% CL)} \\ \noalign{\smallskip}\hline\noalign{\smallskip}
    & Expected [TeV]& Observed [TeV]& Observed (damped) [TeV]\\ \noalign{\smallskip}\hline\noalign{\smallskip}
$n = 2$ & 9.2$^{+0.8}_{-1.0}$ & 7.7$^{+0.4}_{-0.5}$  & 7.7 \\ \noalign{\smallskip}
$n = 3$ & 7.1$^{+0.5}_{-0.6}$ & 6.2$^{+0.4}_{-0.5}$  & 6.2 \\ \noalign{\smallskip}
$n = 4$ & 6.1$^{+0.3}_{-0.4}$ & 5.5$^{+0.3}_{-0.5}$  & 5.5 \\ \noalign{\smallskip}
$n = 5$ & 5.5$^{+0.3}_{-0.3}$ & 5.1$^{+0.3}_{-0.5}$  & 5.1 \\ \noalign{\smallskip}
$n = 6$ & 5.2$^{+0.2}_{-0.3}$ & 4.8$^{+0.3}_{-0.5}$  & 4.8 \\ \noalign{\smallskip}\hline\noalign{\smallskip}
  \end{tabular*}
\end{footnotesize}
\end{center}
\label{tab:add}
\end{table}

% =================

\begin{figure}[htb]
\begin{center}
\includegraphics[width=0.5\textwidth]{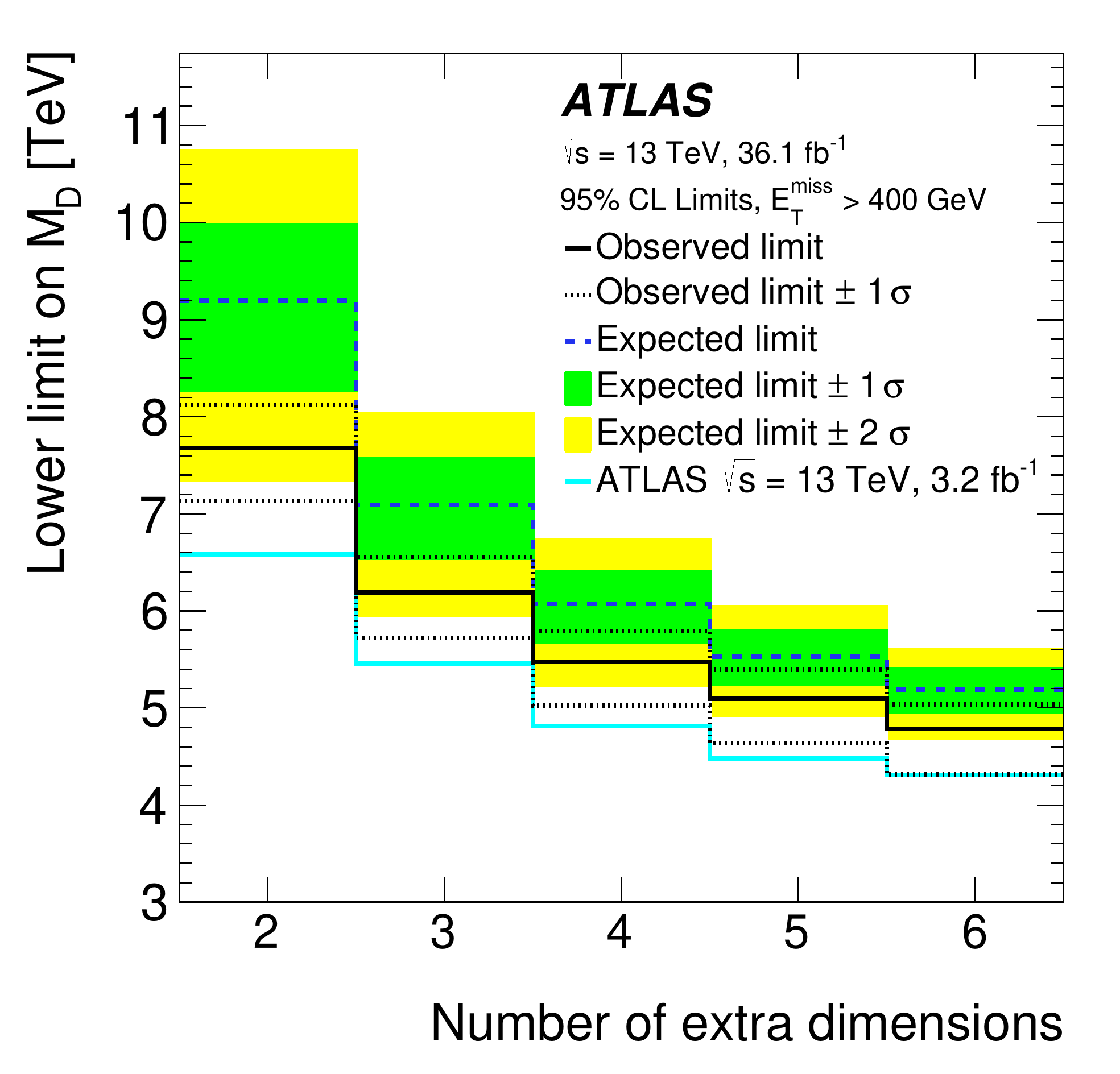}
\end{center}
\caption{
Observed and expected 95$\%$ CL lower limits on the fundamental Planck scale in $4+n$ dimensions, $M_D$, as a function of the number 
of extra dimensions. 
The bands indicate the $\pm 1\sigma$  theory uncertainties in the
observed limit and the $\pm 1\sigma$  and $\pm 2\sigma$ ranges of the expected limit in the absence of a signal.
The 95$\%$ CL limits are computed with no suppression  of the events with 
$\hat{s} > M_D^2$.  
The results from this analysis are compared to previous results from the ATLAS 
Collaboration 
using $3.2 \ifb$ of $\sqrt{s} = 13~\TeV$ data~\cite{Aaboud:2016tnv}.
}
\label{fig:add}
\end{figure}

% =====================
%
% =====================

% All figures and tables should appear before the summary and conclusion.
% The package placeins provides the macro \FloatBarrier to achieve this.
\FloatBarrier

% ====================
% CONCLUSIONS
% ====================
\section{Conclusions}
\label{sec:sum}

Results are reported from a search for new phenomena in events with an 
energetic jet and large missing transverse momentum in proton--proton 
collisions at $\sqrt{s}=13~\TeV$ at the LHC, based on data corresponding 
to an integrated luminosity of $36.1\,\ifb$ collected by the ATLAS 
detector in 2015 and 2016. The measurements are in agreement with the 
SM predictions.  
The results are translated into model-independent 95$\%$~CL   
upper limits on $\sigma \times A \times \epsilon$ in the range 531--1.6~fb, 
decreasing with increasing missing transverse momentum.

The results are translated into exclusion limits on WIMP-pair production. Different simplified models are  
considered  with the exchange of an 
axial-vector, vector  or a pseudoscalar mediator 
in the $s$-channel, and with Dirac fermions as dark-matter candidates.  
In the case of axial-vector or vector mediator models, 
mediator masses below $1.55~\TeV$ are excluded at 95$\%$~CL for very light WIMPs
(for coupling values $g_q = 1/4$ and $g_\chi = 1$), 
whereas the analysis does not have the sensitivity to exclude a pseudoscalar scenario. 
In the case of the axial-vector mediator model, 
the results are translated, in a model-dependent manner, into upper limits on spin-dependent
contributions to the WIMP--nucleon elastic cross section as a function of the WIMP mass.
WIMP--proton cross sections above $2.9 \times 10^{-43}$~cm${}^2$ ($3.5 \times 10^{-43}$~cm${}^2$)
are excluded at 90$\%$ CL for WIMP masses below $10~\GeV$ ($100~\GeV$), complementing results 
from direct-detection experiments. 
In addition, a simplified model of dark-matter production including a coloured scalar mediator   
is considered, for which mediator masses below $1.67~\TeV$ are excluded at 95$\%$~CL for very light WIMPs
(with coupling set to $g = 1$).  

Similarly, the results are interpreted in terms of a search for 
squark-pair production in a compressed-mass supersymmetric scenario. 
In the case of stop- and sbottom-pair production with $\tilde{t}_1 \to  c +  \tilde{\chi}^{0}_{1}$ or 
$\tilde{t}_1 \to  b + ff^{'} + \tilde{\chi}^{0}_{1}$  
and $\tilde{b}_1 \to  b +  \tilde{\chi}^{0}_{1}$, respectively,  squark masses 
below about $430~\GeV$ are excluded at 95$\%$~CL. In the case of squark-pair production with
$\tilde{q} \to  q +  \tilde{\chi}^{0}_{1}  \ (q=u,d,c,s)$,  squark masses below $710~\GeV$ are excluded.  

Finally, the  results are presented 
in terms of lower limits on the fundamental Planck scale $M_D$ in $4+n$ dimensions, versus 
the number of extra spatial dimensions in the ADD LED model.
Values of $M_D$ below $7.7~\TeV$ at $n=2$ and below $4.8~\TeV$ at $n=6$ are excluded at
95$\%$~CL.

%-------------------------------------------------------------------------------
\section*{Acknowledgements}
%-------------------------------------------------------------------------------

% Acknowledgements for papers with collision data
% Version 16-Nov-2017

% Standard acknowledgements start here
%----------------------------------------------
We thank CERN for the very successful operation of the LHC, as well as the
support staff from our institutions without whom ATLAS could not be
operated efficiently.

We acknowledge the support of ANPCyT, Argentina; YerPhI, Armenia; ARC, Australia; BMWFW and FWF, Austria; ANAS, Azerbaijan; SSTC, Belarus; CNPq and FAPESP, Brazil; NSERC, NRC and CFI, Canada; CERN; CONICYT, Chile; CAS, MOST and NSFC, China; COLCIENCIAS, Colombia; MSMT CR, MPO CR and VSC CR, Czech Republic; DNRF and DNSRC, Denmark; IN2P3-CNRS, CEA-DRF/IRFU, France; SRNSF, Georgia; BMBF, HGF, and MPG, Germany; GSRT, Greece; RGC, Hong Kong SAR, China; ISF, I-CORE and Benoziyo Center, Israel; INFN, Italy; MEXT and JSPS, Japan; CNRST, Morocco; NWO, Netherlands; RCN, Norway; MNiSW and NCN, Poland; FCT, Portugal; MNE/IFA, Romania; MES of Russia and NRC KI, Russian Federation; JINR; MESTD, Serbia; MSSR, Slovakia; ARRS and MIZ\v{S}, Slovenia; DST/NRF, South Africa; MINECO, Spain; SRC and Wallenberg Foundation, Sweden; SERI, SNSF and Cantons of Bern and Geneva, Switzerland; MOST, Taiwan; TAEK, Turkey; STFC, United Kingdom; DOE and NSF, United States of America. In addition, individual groups and members have received support from BCKDF, the Canada Council, CANARIE, CRC, Compute Canada, FQRNT, and the Ontario Innovation Trust, Canada; EPLANET, ERC, ERDF, FP7, Horizon 2020 and Marie Sk{\l}odowska-Curie Actions, European Union; Investissements d'Avenir Labex and Idex, ANR, R{\'e}gion Auvergne and Fondation Partager le Savoir, France; DFG and AvH Foundation, Germany; Herakleitos, Thales and Aristeia programmes co-financed by EU-ESF and the Greek NSRF; BSF, GIF and Minerva, Israel; BRF, Norway; CERCA Programme Generalitat de Catalunya, Generalitat Valenciana, Spain; the Royal Society and Leverhulme Trust, United Kingdom.

The crucial computing support from all WLCG partners is acknowledged gratefully, in particular from CERN, the ATLAS Tier-1 facilities at TRIUMF (Canada), NDGF (Denmark, Norway, Sweden), CC-IN2P3 (France), KIT/GridKA (Germany), INFN-CNAF (Italy), NL-T1 (Netherlands), PIC (Spain), ASGC (Taiwan), RAL (UK) and BNL (USA), the Tier-2 facilities worldwide and large non-WLCG resource providers. Major contributors of computing resources are listed in Ref.~\cite{ATL-GEN-PUB-2016-002}.
%----------------------------------------------

%-------------------------------------------------------------------------------
% If you use biblatex and either biber or bibtex to process the bibliography
% just say \printbibliography here
\printbibliography
% If you want to use the traditional BibTeX you need to use the syntax below.
%\bibliographystyle{bib/bst/atlasBibStyleWoTitle}
%\bibliography{mono_paper_2017,bib/ATLAS,bib/CMS,bib/ConfNotes,bib/PubNotes}
%-------------------------------------------------------------------------------

%-------------------------------------------------------------------------------
% Auxiliary material - comment out the following line if you do not have any
% \include{mono_paper_2017-auxmat}
%-------------------------------------------------------------------------------

%-------------------------------------------------------------------------------
% Extra tables etc. for HepData - comment in the following line if you have any
% \include{mono_paper_2017-hepdata}
%-------------------------------------------------------------------------------

%-------------------------------------------------------------------------------
\newpage 
% ATLAS Collaboration author list
% Data extracted on 02-Sep-2017 for paper reference EXOT-2016-27
% \documentclass[11pt]{article}
% \usepackage{a4wide}
% \begin{document}
\begin{flushleft}
{\Large The ATLAS Collaboration}

\bigskip

M.~Aaboud$^\textrm{\scriptsize 137d}$,
G.~Aad$^\textrm{\scriptsize 88}$,
B.~Abbott$^\textrm{\scriptsize 115}$,
O.~Abdinov$^\textrm{\scriptsize 12}$$^{,*}$,
B.~Abeloos$^\textrm{\scriptsize 119}$,
S.H.~Abidi$^\textrm{\scriptsize 161}$,
O.S.~AbouZeid$^\textrm{\scriptsize 139}$,
N.L.~Abraham$^\textrm{\scriptsize 151}$,
H.~Abramowicz$^\textrm{\scriptsize 155}$,
H.~Abreu$^\textrm{\scriptsize 154}$
% ...
\bigskip
\\
% ...
$^{12}$ Institute of Physics, Azerbaijan Academy of Sciences, Baku, Azerbaijan\\
% ...
$^{88}$ CPPM, Aix-Marseille Universit{\'e} and CNRS/IN2P3, Marseille, France\\
% ...
$^{115}$ Homer L. Dodge Department of Physics and Astronomy, University of Oklahoma, Norman OK, United States of America\\
% ...
$^{119}$ LAL, Univ. Paris-Sud, CNRS/IN2P3, Universit{\'e} Paris-Saclay, Orsay, France\\
% ...
$^{137}$ $^{(a)}$ Facult{\'e} des Sciences Ain Chock, R{\'e}seau Universitaire de Physique des Hautes Energies - Universit{\'e} Hassan II, Casablanca; $^{(b)}$ Centre National de l'Energie des Sciences Techniques Nucleaires, Rabat; $^{(c)}$ Facult{\'e} des Sciences Semlalia, Universit{\'e} Cadi Ayyad, LPHEA-Marrakech; $^{(d)}$ Facult{\'e} des Sciences, Universit{\'e} Mohamed Premier and LPTPM, Oujda; $^{(e)}$ Facult{\'e} des sciences, Universit{\'e} Mohammed V, Rabat, Morocco\\
% ...
$^{139}$ Santa Cruz Institute for Particle Physics, University of California Santa Cruz, Santa Cruz CA, United States of America\\
% ...
$^{151}$ Department of Physics and Astronomy, University of Sussex, Brighton, United Kingdom\\
% ...
$^{154}$ Department of Physics, Technion: Israel Institute of Technology, Haifa, Israel\\
$^{155}$ Raymond and Beverly Sackler School of Physics and Astronomy, Tel Aviv University, Tel Aviv, Israel\\
% ...
$^{161}$ Department of Physics, University of Toronto, Toronto ON, Canada\\
% ...
$^{a}$ Also at Department of Physics, King's College London, London, United Kingdom\\
$^{b}$ Also at Institute of Physics, Azerbaijan Academy of Sciences, Baku, Azerbaijan\\
$^{c}$ Also at Novosibirsk State University, Novosibirsk, Russia\\
$^{d}$ Also at TRIUMF, Vancouver BC, Canada\\
$^{e}$ Also at Department of Physics {\&} Astronomy, University of Louisville, Louisville, KY, United States of America\\
% ...
$^{*}$ Deceased
\end{flushleft}

% \end{document}
% Created with xml2latex.py

\end{document}